\documentclass[11 pt,superscriptaddress,longbibliography]{revtex4-1}
\usepackage{textcomp}
\usepackage{times}
\usepackage{graphicx}
\usepackage{float}
\usepackage{latexsym,amsmath,amssymb,bm,euscript}
\usepackage{color}
\usepackage{subfigure}
\usepackage{epstopdf}
\usepackage[colorlinks=true,linkcolor=blue,citecolor=blue]{hyperref}
\usepackage{hyperref}
\usepackage{soul}
\usepackage[normalem]{ulem}
\usepackage{mathrsfs}
\usepackage{amsmath}
\usepackage{lettrine}
\usepackage{xspace}
\usepackage{textcomp}

\newcommand {\B}{\textcolor {blue}}

\def\Th{\ensuremath{T_h}\xspace}
\def\Tl{\ensuremath{T_l}\xspace}

\begin{document}

\title{Kosterlitz-Thouless Melting of Magnetic Order in the Triangular Quantum Ising Material TmMgGaO$_4$}

\author{Han Li}
\affiliation{School of Physics, Key Laboratory of Micro-Nano Measurement-Manipulation and Physics (Ministry of Education), Beihang University, Beijing 100191, China}

\author{Yuan Da Liao}
\affiliation{Beijing National Laboratory for Condensed Matter Physics, and Institute of Physics, Chinese Academy of Sciences, Beijing 100190, China}
\affiliation{School of Physical Sciences, University of Chinese Academy of Sciences, Beijing 100190, China}

\author{Bin-Bin Chen}
\affiliation{School of Physics, Key Laboratory of Micro-Nano Measurement-Manipulation and Physics (Ministry of Education), Beihang University, Beijing 100191, China}
\affiliation{{Munich Center for Quantum Science and Technology (MCQST),
Arnold Sommerfeld Center for Theoretical Physics (ASC) and
Center for NanoScience (CeNS), Ludwig-Maximilians-Universit\"at M\"unchen, Fakult\"at f\"ur Physik, D-80333 M\"unchen, Germany.}}

\author{Xu-Tao Zeng}
\affiliation{School of Physics, Key Laboratory of Micro-Nano Measurement-Manipulation and Physics (Ministry of Education), Beihang University, Beijing 100191, China}

\author{Xian-Lei Sheng}
\affiliation{School of Physics, Key Laboratory of Micro-Nano Measurement-Manipulation and Physics (Ministry of Education), Beihang University, Beijing 100191, China}

\author{Yang Qi}
\email{qiyang@fudan.edu.cn}
\affiliation{Center for Field Theory and Particle Physics, Department of Physics, Fudan University, Shanghai 200433, China}
\affiliation{Collaborative Innovation Center of Advanced Microstructures, Nanjing 210093, China}

\author{Zi Yang Meng}
\email{zymeng@hku.hk}
\affiliation{Beijing National Laboratory for Condensed Matter Physics, and Institute of Physics, Chinese Academy of Sciences, Beijing 100190, China}
\affiliation{Department of Physics and HKU-UCAS Joint Institute of Theoretical and Computational Physics, The University of Hong Kong, Pokfulam Road, Hong Kong, China}
\affiliation{Songshan Lake Materials Laboratory, Dongguan, Guangdong 523808, China}

\author{Wei Li}
\email{w.li@buaa.edu.cn}
\affiliation{School of Physics, Key Laboratory of Micro-Nano Measurement-Manipulation and Physics (Ministry of Education), Beihang University, Beijing 100191, China}
\affiliation{International Research Institute of Multidisciplinary Science, Beihang University, Beijing 100191, China}

\begin{abstract}
\textbf{Frustrated magnets host the promises of material realizations of new paradigm of quantum matter, while direct comparison of unbiased model calculations with experimental measurements is still very challenging. Here, we design and implement a protocol of employing many-body computation methodologies for accurate model calculation -- both equilibrium and dynamical properties -- of a frustrated rare-earth magnet TmMgGaO$_4$ (TMGO), which perfectly explains the corresponding experimental findings. Our results confirm TMGO is an ideal realization of triangular-lattice Ising model with an intrinsic transverse field. The magnetic order of TMGO is predicted to melt through two successive Kosterlitz-Thouless (KT) phase transitions, with a floating KT phase in between. The dynamical spectra calculated suggest remnant images of a vanishing magnetic stripe order that represent vortex-antivortex pairs, resembling rotons in a superfluid helium film. TMGO therefore constitutes a rare quantum magnet for realizing KT physics and we further propose experimental detections of its intriguing properties.}
\end{abstract}

\date{\today}
\maketitle

\lettrine[lines=2, findent=3pt, nindent=0pt]{K}osterlitz-Thouless (KT) physics bestows interesting mechanism of phase transition upon two-dimension (2D) interacting system with a continuous symmetry. Although such symmetry is not allowed to break spontaneously at any finite temperature~\cite{Mermin-Wagner}, phase transition can still take place from the high temperature disordered phase to a KT phase with quasi-long-range order, which has a topological root in the binding of the vortex and antivortex pair~\cite{Kosterlitz1973,Kosterlitz1974}. Experimentally, the KT transition has been observed in thin helium films \cite{Bishop1978} and ultracold 2D bose gases \cite{Hadzibabic2006,Desbuquois2012}. Two distinct types of elementary excitations, i.e., phonons and rotons, play essential roles in the related superfluid phenomena \cite{Landau1949,Feynman1955}, and they are important for the understanding of liquid helium thermodynamics~\cite{Bendt1959}. Besides interacting bosons in liquid and gas, there are also theoretical proposals of KT transitions in solid-state magnetic systems such as the 2D classical XY ~\cite{Kosterlitz1973,Kosterlitz1974} and the frustrated quantum Ising models~\cite{Isakov2003}. However, to date, the material realization of the KT transition in 2D magnets has rarely been reported.

\begin{figure*}[t!]
   \includegraphics[angle=0,width=0.99\linewidth]{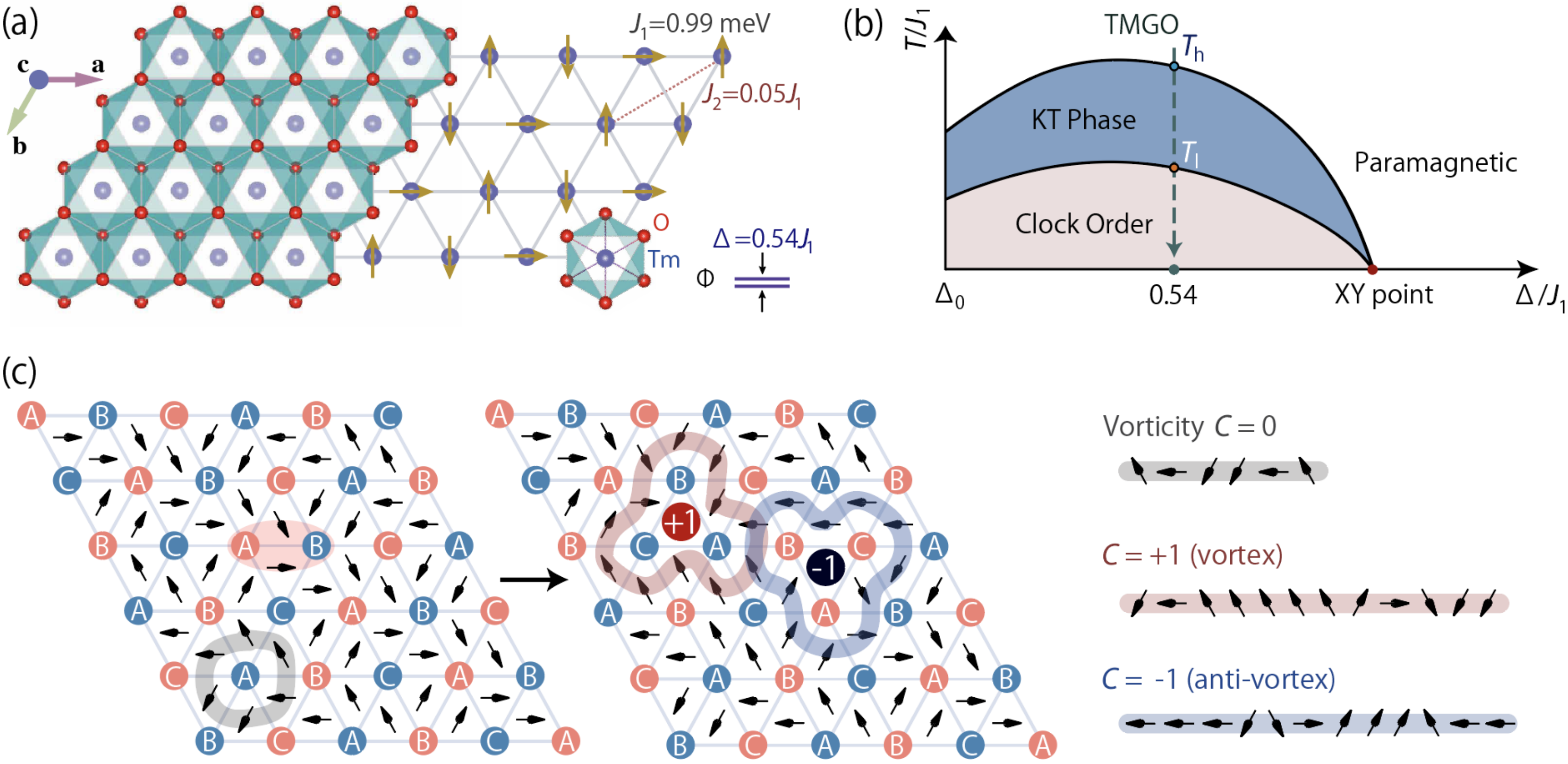}
    \caption{\textbf{The crystal structure, phase diagram, and spin texture in quantum magnet TMGO.}  (a) The Tm$^{3+}$ ions, with an energy splitting $\Delta$ between two lowest non-Kramers levels $|\Phi_{\pm}\rangle$, constitute an effective spin-1/2 model on the triangular lattice, with $J_1$ and $J_2$ interactions. An illustration of spin structure in the clock phase is provided, where the spin-up and spin-down arrows are along the magnetic easy $c$-axis, and the horizontal arrow stands for superposition of spin up and down, i.e., $|\rightarrow\rangle$. (b) The schematic phase diagram of quantum TLI model, there exists a quantum critical point on the horizontal axis with emergent spin XY [i.e., U(1)] symmetry, which extends into an intriguing KT phase at finite $T$. $\Delta> \Delta_0$ stands for transverse fields where the clock order is stabilized in the ground state, in the $J_1$-$J_2$ model with small $J_2$ coupling. The vertical arrowed line along $\Delta=0.54J_1$ represents the TMGO material, with two KT transitions at $T_{\rm{h}}$ and $\Tl$, respectively. (c) shows the magnetic stripe order, with the red sites for spin up ($m^z=1/2$) and blue ones spin down ($m^z=-1/2$) on three sublattices A, B, and C. The pseudo spins, i.e., complex order parameters $\psi$ in Eq.~(\ref{Eq:psi}), are plotted as arrows rotating within the plane, and a vortex-antivortex pair is created by flipping simultaneously two spins within the red oval in the left subpanel. As tracked along the paths (exemplary paths are indicated in the plot), topological charge $C=\pm 1$, corresponding to $2\pi$ clockwise/counterclockwise angle, emerges when the pseudo spins wind clockwise around the vortex/antivortex, and zero vorticity appears when counting the pseudo spin winding of both (or no defect at all).}
\label{Fig:Illustrator}
\end{figure*}

In the mean time, the search of exotic quantum magnetic states in the triangular lattice spin systems --  
the motif of frustrated magnets -- has attracted great attention over the decades.
Experimentally, the triangular lattice quantum magnets have been synthesized only very recently, including
compounds Ba$_3$CoSb$_2$O$_9$ \cite{Zhou2012,Susuki2013,Ma2016,Ito2017}, Ba$_8$CoNb$_6$O$_{24}$ \cite{Rawl2017,Cui2018}, 
and a rare earth oxide YbMgGaO$_4$ -- which has been suggested as a quantum spin liquid candidate~\cite{Li2015b,Li2016,Shen2016,Li2019}, 
while an alternative scenario of glassy and disordered induced state has also been proposed recently~\cite{Zhu2017,Kimchi2018,JSWen2018}. 
On the other hand, an Ising-type triangular antiferromagnet TmMgGaO$_4$ (TMGO, with Yb$^{3+}$ replaced by another rare earth ion 
Tm$^{3+}$)~\cite{Cava2018,Li2018,Shen2018}, as shown in Fig.~\ref{Fig:Illustrator}(a) and explained in details in this paper, is the successful material realization of a quantum magnet with strong Ising anisotropy.

In this work, we construct the microscopic model of TMGO and employ two state-of-the-art quantum many-body simulation approaches: 
the exponential tensor renormalization group (XTRG)~\cite{Chen2018} and 
quantum Monte Carlo equipped with stochastic analytic
continuation (QMC-SAC)~\cite{Sandvik2016,YQQin2017,Shao2017,CJHuang2018,GYSun2018}, to calculate both the thermodynamic and
dynamic properties. By scanning various parameters and fit our simulation results to the existing experimental data~\cite{Cava2018,Li2018,Shen2018},
we find TMGO realizes a triangular-lattice transverse field Ising model and determine accurately its model parameters. 
Based on this, we conclude that TMGO should host the celebrated KT phase, 
and further predict several prominent features to be observed in TMGO, 
inspired by the experimental measurements for detecting KT physics in a superfluid thin film \cite{Bishop1978}. 
It is worthwhile to point out that our calculation of quantum fluctuations goes beyond the linear spin-wave approximation 
in Ref.~\cite{Shen2018} and puts the system in clock ordered (later melted through KT transitions), 
rather than disordered, regime. Therefore, our methods and results do not only explain the experimental findings,
but more importantly establish a protocol of acquiring equilibrium and dynamic experiments of strongly correlated quantum material, 
such as TMGO, in an unbiased manner.

\section*{Results}
\textbf{Microscopic spin model.} Due to strong spin-orbit coupling and crystal electric field splitting, TMGO can be described as an effective spin-1/2 model with strong easy-axis anisotropy, i.e., a triangular lattice Ising model (TLI), as shown in Fig.~\ref{Fig:Illustrator}(a). First-principle calculations of the TMGO material are performed based on the density functional theory (DFT) \cite{VASP1,VASP2}, where we see a large easy-axis anisotropy of room-temperature energy scale. By comparing the DFT energies of antiferromagnetic spin configuration to a ferromagnetic one, one finds the former has a lower energy, and the coupling strength can be estimated as a few tenths of meV. Density distribution of 4f electrons in TMGO can also be obtained, where it is observed that the 4f electrons of Tm$^{3+}$ are coupled via superexchange mediated by 2p electrons of O$^{2-}$ within the triangular lattice plane (Supplementary Note~1).

In Ref.~\onlinecite{Li2018}, the authors took the lowest two levels $|\Phi_{\pm} \rangle$ of Tm$^{3+}$ 
as non-Kramers doublet, and construct a classical TLI with both nearest-neighbor (NN) and next-nearest-neighbor (NNN) 
interactions to account for the absence of zero point entropy observed in experiments. 
Substantial randomness was also introduced to explain the smooth magnetization curves even at very low temperature. 
Later on, inelastic neutron scattering (INS) results of TMGO reveal a clear magnon band~\cite{Shen2018}, 
suggesting the influence of coupling randomness should be modest in TMGO,
and an adequate modelling of the material shall include in the Hamiltonian non-commuting terms with quantum fluctuations. 
Since the Kramers theorem is absent in Tm$^{3+}$ system with total angular momentum $J=6$, 
a small level splitting $\Delta$ between the quasi-doublet $|\Phi_{\pm} \rangle$  is involved, 
as shown in Fig.~\ref{Fig:Illustrator}(a). Therefore, a quantum TLI model was proposed \cite{Shen2018}, with spin-1/2 Hamiltonian
\begin{equation}
H_{\rm{TLI}} = J_1\sum_{\langle i,j \rangle} S_i^z S_j^z + J_2 \sum_{\langle\langle i,j \rangle\rangle} S_i^z S_j^z \nonumber - \sum_i (\Delta S_i^x + h \, g_{\parallel} \mu_B \, S_i^z),
\label{Eq:qTLI}
\end{equation}
where $\langle , \rangle$ $\big(\langle\langle , \rangle\rangle \big)$ stands for NN (NNN) couplings $J_1$ ($J_2$), $\Delta$ the energy splitting between $|\Phi_{\pm} \rangle$ (i.e., the intrinsic transverse field), and $h$ is the external magnetic field. $g_{\parallel} = 2 J g_J$ constitutes the effective spin-1/2 $g$ factor, 
with $g_J$ the  Land\'e factor. 

The phase diagram of quantum TLI has been studied intensively with analytic and numeric methods in the 
past~\cite{Isakov2003,YCWang2017,Powalski2013,Fey2019} and is schematically shown in Fig.~\ref{Fig:Illustrator}(b). 
We indicate the TMGO model parameter with the vertical arrow (the determination of parameters is given below). 
From high to low temperatures, the system first goes through a KT transition at \Th from the paramagnetic phase to a KT phase with 
power-law (algebraic) spin correlations. At a lower temperature $\Tl$, the system  enters the clock phase with a true long-range order 
depicted in Fig.~\ref{Fig:Illustrator}(a). This three-sublattice clock order breaks the discrete lattice point group as well as the $Z_2$ 
spin symmetries, giving rise to a low but finite transition temperature $\Tl$.

Increasing the next-nearest neighbor couplings $J_2$, say, at $ J_2/J_1=0.2$,  we find the static magnetic structure factor develops a stripe order \cite{Metcalf1974} with structure factor peak at M point of the Brillouin zone (BZ) (see Supplementary  Notes~2,3). This magnetic stripe order, as shown in Fig.~\ref{Fig:Illustrator}(c), has been observed previously in TLI material AgNiO$_2$, where $J_2$ coupling is relatively strong ($\sim 0.15 J_1$, along with other interactions) and the exotic KT physics is absent there \cite{Wheeler2009}. In TMGO, however, the clock order wins over the stripe order as $J_2/J_1 \simeq 0.05$ is relatively small in this material. Nevertheless, as will be shown below, a ghost of the latter order -- the M rotonlike modes -- remains in the spin spectrum \cite{Shen2018}, which turns out to be related to a vortex-antivortex pair excitation in the topological language (see Discussion section). 

\begin{figure*}[t!]
\includegraphics[angle=0,width=0.8\linewidth]{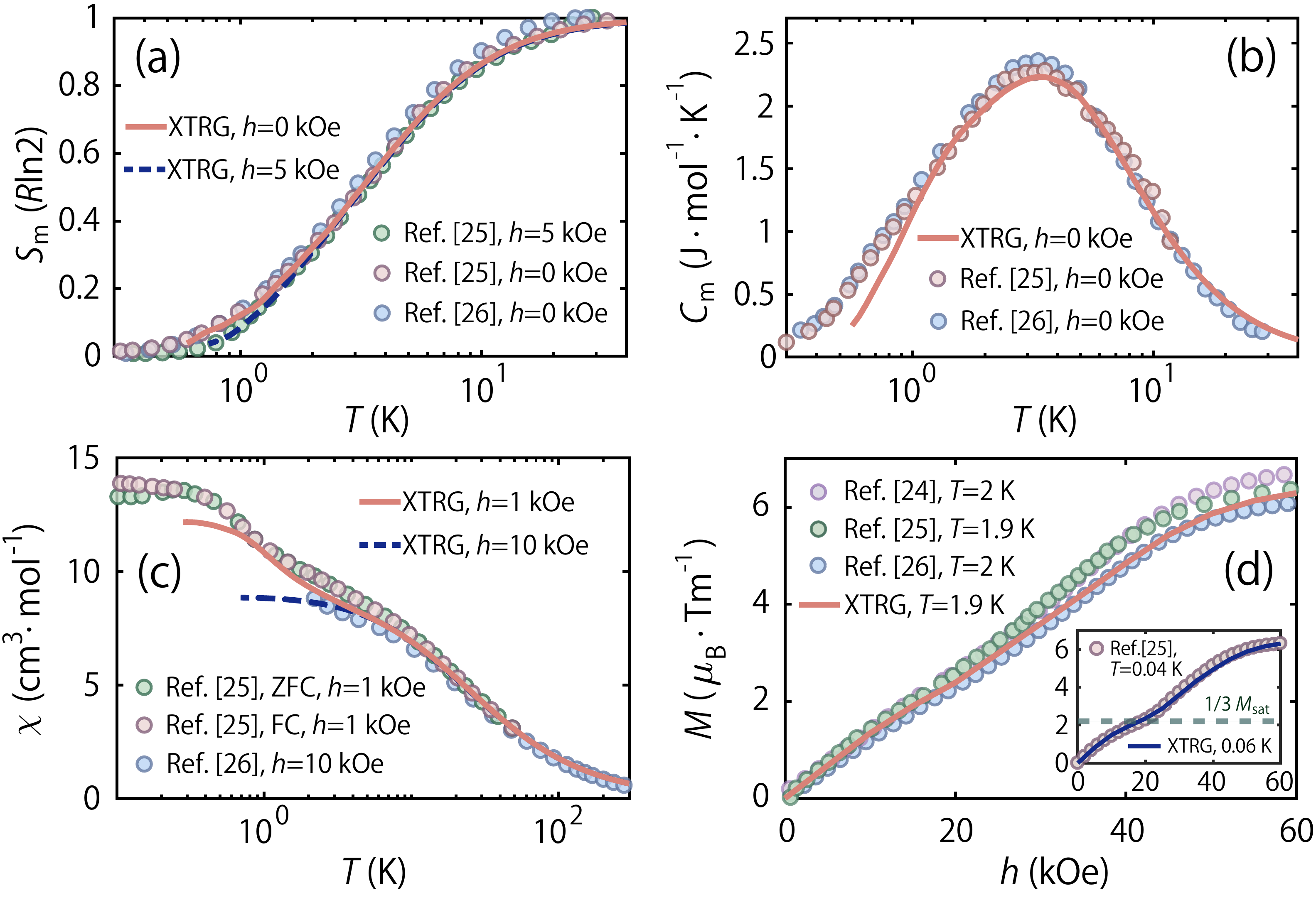}
\caption{\textbf{Thermodynamic measurements and XTRG fittings to experimental results.} Agreements between the experimental curves (taken from various independent measurements \cite{Cava2018,Li2018,Shen2018}) and numerical results can be seen in all panels, down to very low temperatures. (a) includes two experimental entropy curves under fields $h=0$ and $5$ kOe~\cite{Li2018,Shen2018}, and the specific heat data shown in (b) are under zero field. The magnetic susceptibility $\chi$ is shown in panel (c), which follows firstly the Curie-Weiss law at high temperature, i.e., $\chi \sim C/(T-\Theta$) with $\Theta\simeq -19$ K (see Supplementary Note 5), and then exhibits at $\sim 10$ K a shoulder structure, signifying the onset of antiferromagnetic correlation. For $T \lesssim \Th^* \sim$ 4 K, $\chi$ rises up again and eventually converges to a finite value as $T$ decreases to below $\Tl^* \sim 1$ K. These anomalous susceptibility curves can be fitted very well by our  simulations and naturally understood within the framework of TLI model. For the magnetization curves in (d), the perfect consistency between numerics and experiments hold for both intermediate-$T$ ($\simeq 2$ K) and the low-$T$ ($40$ - $60$ mK) curves. The latter is shown in the inset, where a quasi-plateau at $M \simeq \frac{1}{3} M_{\rm{sat}}$ becomes prominent, with  $M_{\rm{sat}}=Jg_J \mu_B$ the saturation magnetization of TMGO.}
\label{Fig:Thermo}
\end{figure*}

\begin{figure*}[t!]
\includegraphics[angle=0,width=0.6\linewidth]{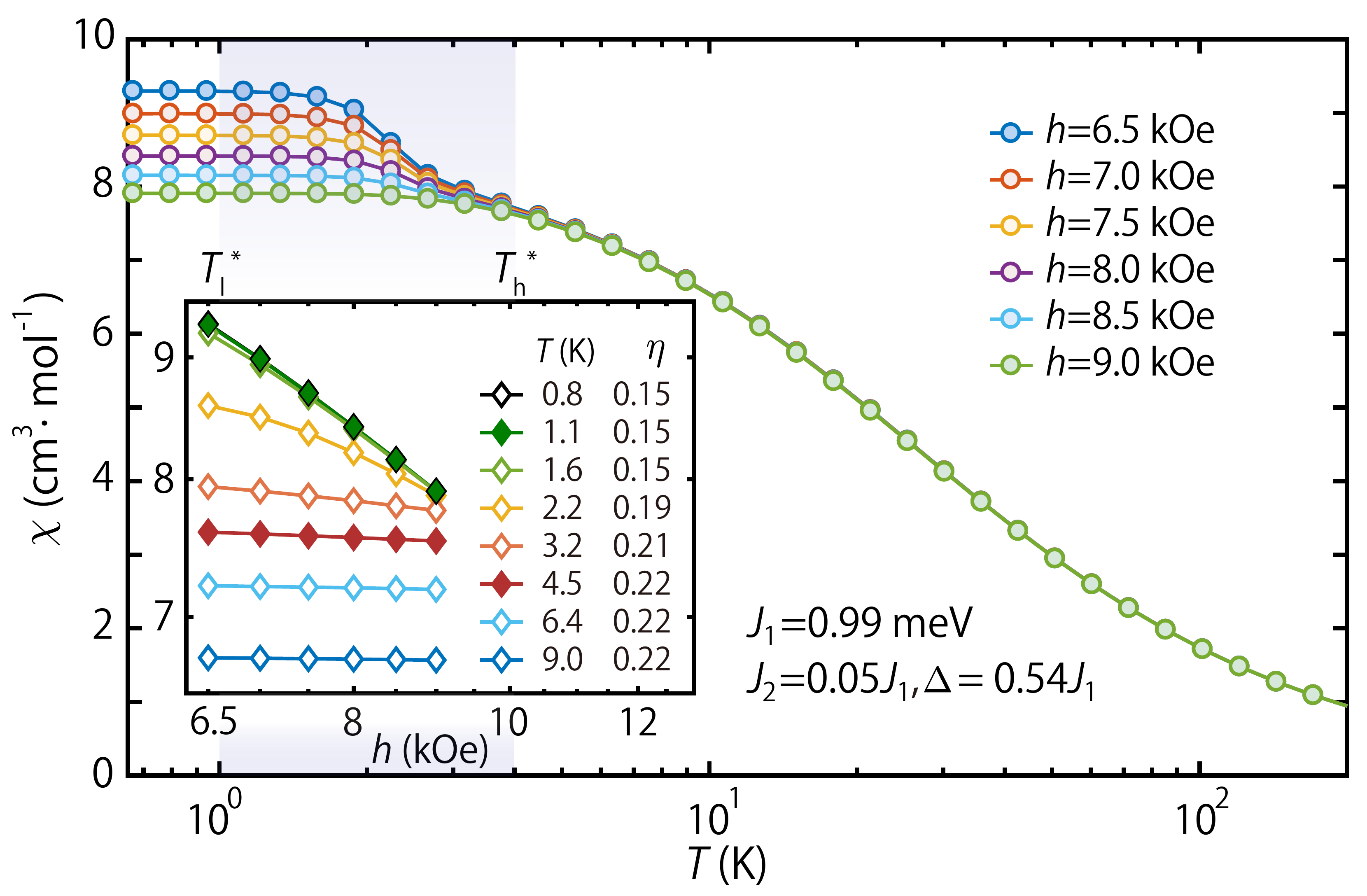}
  \caption{\textbf{Algebraic scaling of uniform magnetic susceptibility.}  At high temperature $T\gtrsim4$ K, $\chi(h,T)$ remains independent of $h$ for small fields, which indicates an exponent of $\alpha=0$ in $\chi(h,T) \sim h^{-\alpha(T)}$, while for $1$ K $ \lesssim T  \lesssim$ $4$ K (shaded regime), $\chi(h,T)$ exhibits an universal scaling vs. $h$ with $\alpha \neq 0$. For temperatures below $\Tl^* \sim 1$ K, the susceptibility ceases to increase as the system orders into the clock phase. In the inset, $\chi(h,T)$ vs. $h$ curves are presented in the log-log plot, where the algebraic scaling manifests itself in the KT phase (and in the clock phase due to the saturation). The fitted $\eta$ values decrease from $2/9$ gradually to $\sim 0.15$, the deviation of the latter from the expectation of $1/9$ is ascribed to limited lattice size ($L=12$ torus) adopted in the QMC calculations.}
\label{Fig:Sus}
\end{figure*}

 \textbf{Thermodynamics and parameter fittings.} The model parameters in Eq.~(\ref{Eq:qTLI}) can be accurately determined through fitting the available experimental data of TMGO~\cite{Cava2018,Li2018,Shen2018}, from which we find $J_1=0.99$ meV, $J_2=0.05J_1$, $\Delta=0.54 J_1$ and $g_{\parallel} = 13.212$. We present in Fig.~\ref{Fig:Thermo} the calculated thermodynamic quantities and their experimental counterparts \cite{Li2018,Shen2018}, where excellent agreements are seen. In Fig.~\ref{Fig:Thermo}(a), at high temperatures $T>30$~K, the magnetic entropy $S_{\rm{m}}$ approaches $R \ln{2}$, corresponding to the paramagnetic phase with effective spin-1/2. As temperature decreases, $S_{\rm{m}}$ gradually releases throughout the intermediate-temperature regime and approaches zero below $\Tl^* \simeq 1$~K, as the long-range clock order develops. In Fig.~\ref{Fig:Thermo}(b) the very good agreement in magnetic specific heat $C_{\rm{m}}$ extends from high $T$ ($\sim 30$ K) all the way down to low temperature $\sim$1 K.

In the fittings of the magnetic entropy and specific heat curves in Figs.~\ref{Fig:Thermo}(a,b), we rescale the $T$-axes (in the unit of $J_1$) to lay the model calculations on top of the experimental measurements, and in this way we find the optimized $J_1 =$ 0.99 meV. In both plots, the $y$-axes scaling ratios are associated with the ideal gas constant $R=8.313$ J$\cdot$mol$^{-1} \cdot$K$^{-1}$ and thus fixed. In Fig.~\ref{Fig:Thermo}(c), we fit the magnetic susceptibility $\chi(h) = \frac{M(h)}{h}$, with $M(h)$ the uniform magnetization (per Tm$^{3+}$), under external fields of a small value $h=1$ kOe and a larger one $10$ kOe. As shown in Fig.~\ref{Fig:Thermo}(c), by setting the effective Land\'e factor $g_{\parallel} = 13.212$, we fit both susceptibility curves very well. This completes the model parameters in the Hamiltonian Eq.~(\ref{Eq:qTLI}) (see more fitting details in Supplementary Note 4), and note this parameter set also leads to accurate entropy results at 5 kOe when put in a direct comparison to the experimental line in Fig.~\ref{Fig:Thermo}(a).  
     
With the parameters $J_1$, $J_2$, $\Delta$, and $g_\parallel$ determined from above fittings, we can compute the magnetization curve $M(h)$ and compare it directly with the independently measured experimental curves in Fig.~\ref{Fig:Thermo}(d). Note there exists a turning point at about 1/3 magnetization, under a magnetic field around 20-25 kOe. It becomes clearer as $T$ decreases further down to 40 mK [see the inset of Fig.~2(d)], suggesting the existence of field-induced quantum phase transition in TMGO. This sharp change of behaviors can also be witnessed in the specific heat curves under various magnetic fields (see Supplementary Note~6), which can also be understood very well within the set of parameters obtained above.

Lastly, we briefly discuss the scalings in the uniform susceptibility $\chi$ in Fig.~\ref{Fig:Thermo}(c) as well as Fig.~\ref{Fig:Sus}, where susceptibility curves under more external fields are computed theoretically in the later plot, as a complementary. These $\chi$ data reflects the two-step establishment of magnetic order as $T$ lowers. As pointed out in Refs.~\onlinecite{Damle2015,Biswas2018}, a universal scaling $\chi(h) = h^{-\alpha}$ appears for small fields $h$ with $\alpha=\frac{4-18\eta(T)}{4-9\eta(T)}$ in the KT regime, where $\eta(T) \in [\frac{1}{9},\frac{2}{9}]$ is the anomalous dimension exponent of the emergent XY order parameter varying with temperature. For $T=\Tl$, $\eta=\frac{1}{9}$ and $\chi(h) \sim h^{-2/3}$, which diverges as $h$ approaches zero; while for some higher temperature (slightly) below \Th, $\eta=\frac{2}{9}$ and $\chi(h)$ remains a constant vs. $h$. Therefore, at small external field, the increase of $\chi$ at intermediate $T$ reflects the decrease of $\eta(T)$ vs. $T$, and such enhancement becomes less prominent for a relatively larger field, say, $h=10$ kOe. This salient difference indeed can be noticed in the experimental as well as our numerical curves in Fig.~\ref{Fig:Thermo}(c). Moreover, in Fig.~\ref{Fig:Sus} we calculate more susceptibility curves under various magnetic fields $h$ between 6.5 kOe and 9 kOe, where the power-law scaling is shown explicitly and the anomalous exponent $\eta$ can be extracted therein.

\begin{figure*}[t!]
\includegraphics[angle=0,width=0.8\linewidth]{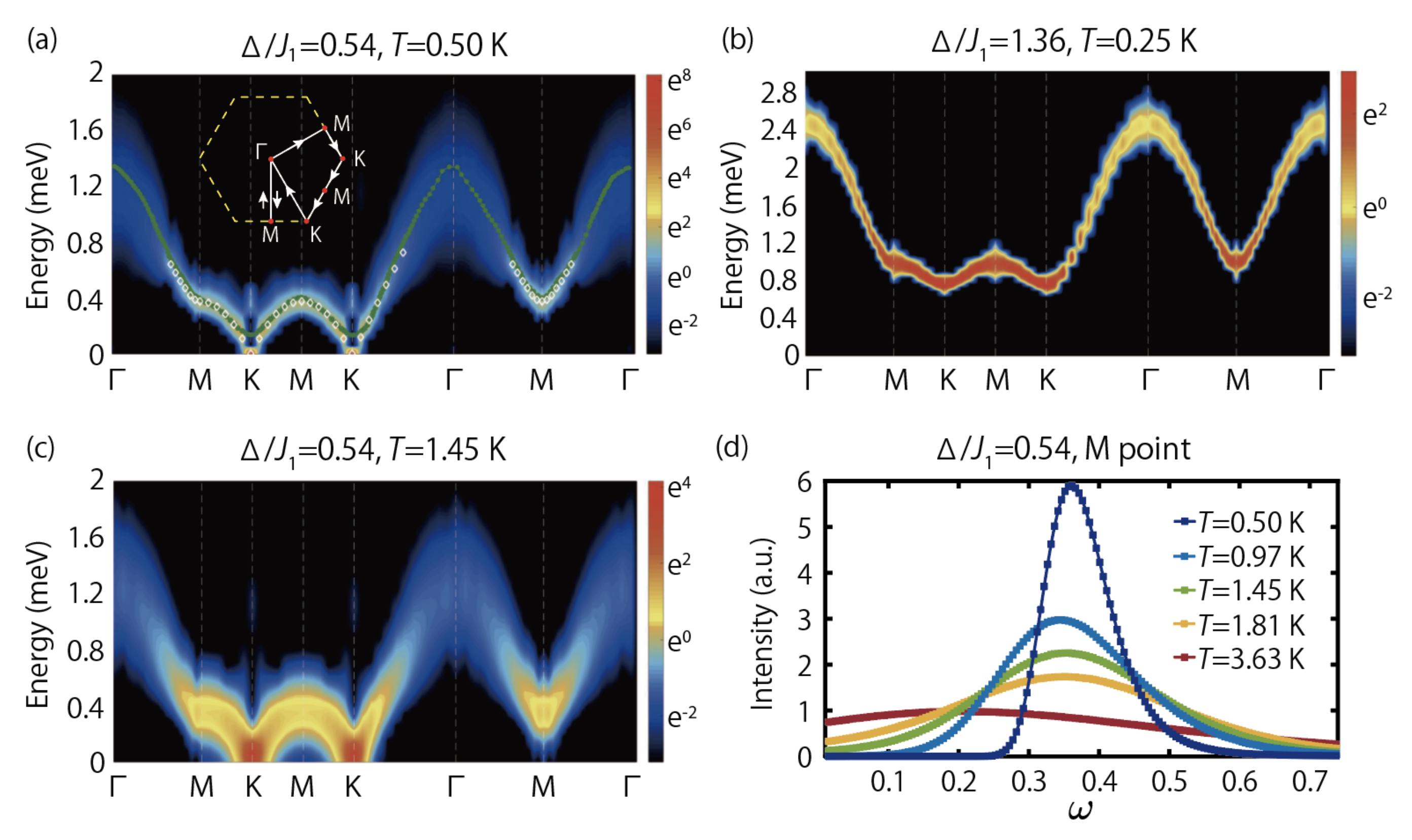}
\caption{\textbf{QMC-SAC spectra of TMGO at finite temperature.} (a) shows the calculated spin spectrum of TLI at $T=0.5$ K, with model parameters determined in this work and computed on an $L=36$ torus geometry. The computed spectrum agrees excellently with the experimental INS results taken from Ref.~\cite{Shen2018}, whose peak positions are shown as the green dotted line. The white diamonds represent the bottom-part excitation results without resorting to analytical continuation, which agree very well with the QMC-SAC results. The path consists of the $\Gamma$-M-K-M-K-$\Gamma$ loop and a $\Gamma$-M-$\Gamma$ vertical mini-loop, as shown in the inset. (b) Spectrum with the parameter set given in Ref.~\cite{Shen2018}, which clearly fails to describe the material. In (c) we plot in the spin spectra with model parameters in (a) but at a higher temperature 1.45 K. Compared to (a), the K point gap gets smoothed and the rotonlike gap reduced. We collect the M point intensity vs. $\omega$ and plot in (d), where the linewidth near the rotonlike minima is substantially broadened as $T$ increases, suggesting strong fluctuations and vortex proliferation in the system.
}
\label{Fig:Dynamics}
\end{figure*}

\textbf{Spin spectra and magnetic structure factors.}
Spin frustration can lead to strong renormalization effects, which in turn gives rise to interesting spectrum. Here we employ the QMC-SAC approach~\cite{Sandvik2016,YQQin2017,Shao2017,CJHuang2018,GYSun2018} to compute the spin spectra $S(\mathbf{q},\omega)$ from  $S^z$ spin correlations, at various temperatures (see Methods). The obtained spectra, with model parameters determined from equilibrium data fittings, are plotted in Fig.~\ref{Fig:Dynamics} and compared directly to the INS results \cite{Shen2018}. Figure~\ref{Fig:Dynamics}(a) depicts the spin spectrum inside the clock phase, at a low temperature $T = 0.5$ K. Since the clock phase is of discrete symmetry breaking, the $S($$\mathbf{q}$$,\omega=0)$ signals the Bragg peak of the clock order and there exists a small gap $\sim 0.1$ meV between the $\omega=0$ and finite $\omega$ spectra, consistent with the INS result. The rotonlike modes are also clearly present in the QMC-SAC results, with an energy gap $\simeq 0.4$ meV, in quantitative agreement with that in Ref.~\cite{Shen2018}. 

Figure~\ref{Fig:Dynamics}(b) is the QMC-SAC spectrum calculated according to the parameters (${\Delta}/J_1\simeq1.36$, $J_2/J_1 \simeq 0.046$) given in Ref.~\cite{Shen2018}. As mentioned in the introduction, we find, via spin structure factor calculations, that such set of parameters actually put the model in the disordered paramagnetic phase with $\Delta > \Delta_{\rm{c}} \sim 0.8 J_1$. It is possible that the fitting scheme adopted in Ref.~\cite{Shen2018} is based on mean-field treatment and cannot capture the quantum fluctuations inherent to the quantum TLI model as well as the material TMGO. This is a clear sign that the unbiased quantum many-body calculation scheme in our work is the adequate approach to explain the experimental results.

We continue with the correct parameter set and rise the temperature to $T=1.45$ K in Fig.~\ref{Fig:Dynamics}(c). It is interesting to see that the dispersion still resembles that in the clock phase of Fig.~\ref{Fig:Dynamics}(a) but with a vanishing gap at the K point, as well as softened M roton modes. To show it more clearly, we plot the intensity at M in Fig.~\ref{Fig:Dynamics}(d), where the roton gap gets reduced as $T$ increases, with substantially broadened linewidths.  Since M rotonlike excitations can be related to vortex-pair excitation (see Discussions), this softening of M roton is consistent with the scenario of vortex proliferations near the upper KT transition. Such remarkable spectra constitute a nontrivial prediction to be confirmed in future INS experiments.

\begin{figure*}[t!]
\includegraphics[angle=0,width=0.75\linewidth]{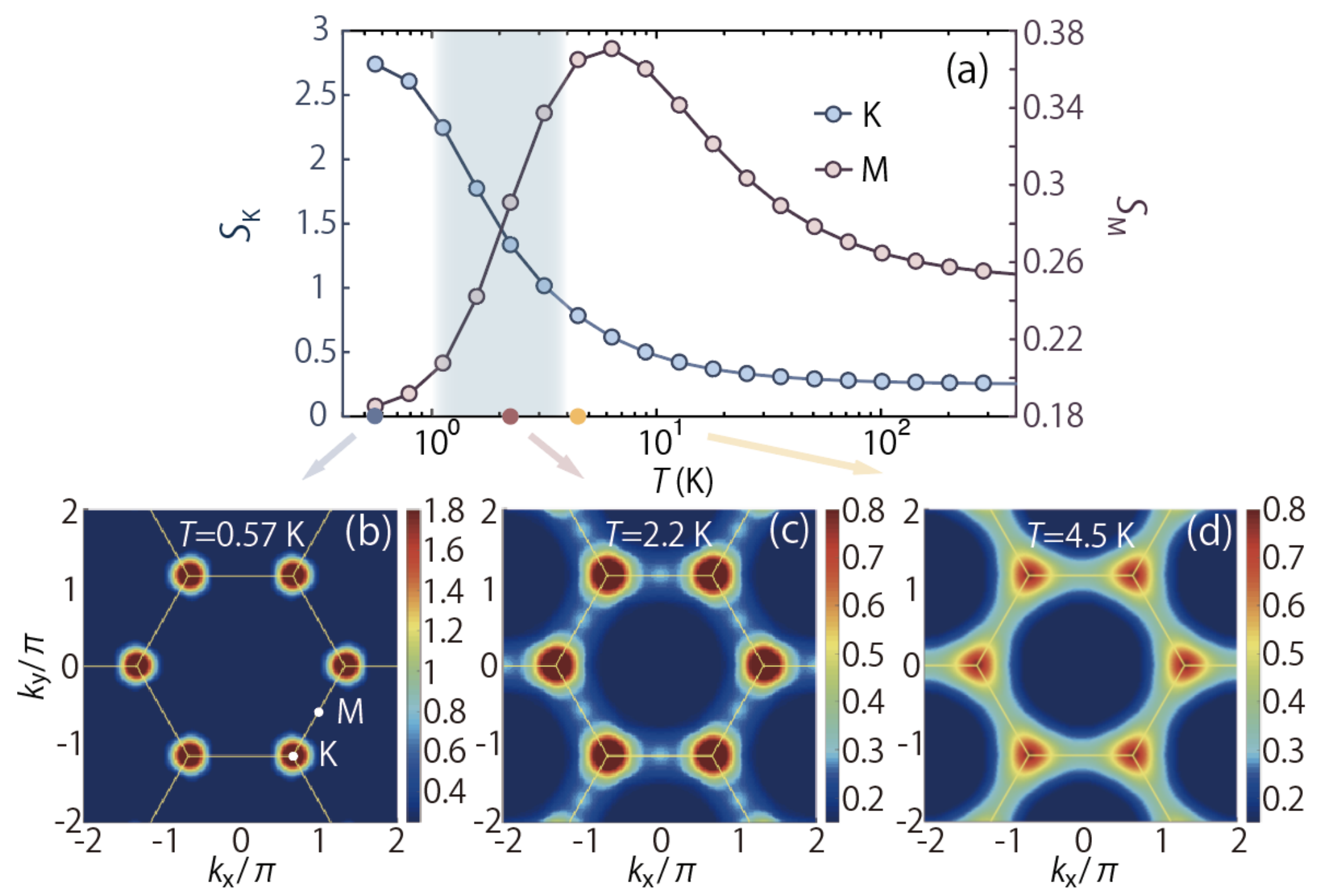}
  \caption{\textbf{Static magnetic structure factor of TMGO at finite temperatures.} (a) The temperature dependence of the structure factor $S(\mathbf{q})$ at M and K points indicated in panel (b). The $S_\textup{M}$ value enhances anomalously in the intermediate temperatures, while the K-point peaks become significantly strengthened, representing the full establishment of the three-sublattice clock order at $T=0.57$ K as shown in (b). Panel (c) shows the $S(\mathbf{q})$ contour at $T=2.2$ K, where a ``ghost" peak appears at M, representing short-range stripy correlation. In panel (d), as $T$ further enhances to 4.5 K, the K peak clearly weakens and spreads out, and the M ghost peak becomes virtually invisible.}
\label{Fig:Sq}
\end{figure*}

Besides, the static magnetic structure factor $S(\mathbf{q}) = \sum_{i,j}$ e$^{i \mathbf{q}\cdot(\mathbf{r}_i-\mathbf{r}_j)} \langle S_{i}^z \, S_j^z \rangle$ are also simulated, where $\mathbf{r}_{i}$ and $\mathbf{r}_j$ run throughout the lattice. Fig.~\ref{Fig:Sq}(a) shows the temperature dependence of $S_\textup{K}$ and $S_\textup{M}$, where one observes an enhancement of $S_\textup{M}$ at intermediate temperature, signifying its closeness towards the stripe order. At $T<\Tl^*$, the enhancement of $S_\textup{M}$ vanishes and instead the $S_\textup{K}$ intensity becomes fully dominant.  Figures~\ref{Fig:Sq} (b,c,d) show the $S(\mathbf{q})$ results at low ($T=0.57$ K), intermediate ($T= 2.2$ K), and high temperaure ($T=4.5$ K). In the clock phase, $S(\mathbf{q})$ evidently peaks at the K point, the ordering wavevector of the three-sublattice clock phase, while in the intermediate temperature regime, notably there exists a ``ghost" peak at the M point, manifesting the existence of short-range stripe order selected by thermal fluctuations. These interesting features are gone at higher $T=4.5$ K, where strong fluctuations considerably weaken the structure factor peaks.

\section*{Discussion}
As postulated by Landau, the phonon-roton spectrum play an important role in understanding low-$T$ thermodynamics and critical velocities of superfluidity, etc, in liquid helium, where the rotons are believed to be related to local vorticity of the fluid \cite{Landau1949}. Roton constitutes a minima in the spectrum at finite momentum and energy, and has been regarded as a quantum analog of hydrodynamic vortex ring, as coined ``the ghost of vanished vortex ring" by Onsager~\cite{Donnelly1974}. On the other hand, as derived from a trial wavefunction introduced by Feynman \cite{Feynman1955}, the roton excitation has energy $\epsilon_\mathbf{q} \sim \hbar \mathbf{q}^2/[2m \, S(\mathbf{q})]$, with $m$ is the helium atom mass and $S(\mathbf{q})$ the structure factor peak, and it is therefore also associated with an incipient crystalization Bragg peak competing with superfluidity~\cite{Nozieres2004}. Rotonlike excitations are also found in thin helium films \cite{Carneiro1976} and frustrated triangular Heisenberg (TLH) magnets. In the latter case, M point rotonlike modes were predicted theoretically \cite{Starykh2006,Chernyshev06,Zheng2006PRL,Zheng2006PRB} and confirmed in recent experiments \cite{Ma2016, Ito2017}, whose nature is under ongoing investigations~\cite{Verresen2019,Farrari2019}. Notably, it has been proposed that the M rotons in TLH can be softened by further enhancing spin frustration (and thus quantum fluctuations) \cite{Farrari2019} or thermal fluctuations \cite{Chen2019}, which melts the long-range or incipient semi-classical 120$^\circ$ order, driving the system into liquid-like spin states.

In the frustrated magnet TMGO, as mentioned earlier in Fig.~\ref{Fig:Dynamics}, there exists rotonlike modes with large density of states, which becomes softened even at low temperatures, melting the clock order, and strongly influences thermodynamics of the system. Like in liquid helium, roton in TMGO has also a topological origin: we demonstrate below that the rotonlike modes represent bound states of topological vortex excitations, via a pseudo-spin mapping of the spin stripe order. As shown in Fig.~\ref{Fig:Illustrator}(c), the stripe order constitutes a proximate competing order to the clock state. Although it eventually gets perished in TMGO at low $T$, the stripe order leaves a ``ghost image" in the excitation spectrum, i.e., rotonlike dip along the $\Gamma$-M-$\Gamma$ path  in Fig.~\ref{Fig:Dynamics}(a). Correspondingly, there exists an incipient $S_\textup{M}$ peak in the static structure factor at intermediate $T$ [Fig.~\ref{Fig:Sq}(c)], i.e., $\epsilon_\mathbf{q} \sim 1/S(\mathbf{q})$, similar to rotons in the superfluid helium discussed above \cite{Feynman1955,Nozieres2004}. 

On top of the spin stripe order, the M rotonlike excitation can be related to a locally bounded vortex pair, 
some form of ``rotational motion" happening in TMGO. We perform a pseudo-spin mapping
\begin{equation}
\psi = m^z_{\rm A} + {\rm e}^{i2\pi/3} \, m^z_{\rm B}  +  {\rm e}^{i 4\pi/3} \, m_{\rm C}^z,
\label{Eq:psi}
\end{equation}
where $\psi = |\psi| {\rm e}^{i \theta}$ is the complex order parameter \cite{YCWang2017}, i.e., the pseudo spin. As shown in Fig.~\ref{Fig:Illustrator}(c), $\psi$ is located in the center of each triangle, with emergent XY degree of freedom $\theta$. In Eq.~(\ref{Eq:psi}), $m_\gamma^z  = \pm 1/2$ represents the spin-up(-down) of corresponding spin $S^z$ components at $\gamma$-sublattice ($\gamma = {\rm A,B,C}$). This mapping helps establishing a Landau-Ginzburg theory of TLI \cite{Isakov2003,YCWang2017}, and the clock order shown in Fig.~\ref{Fig:Illustrator}(a) corresponds to a ferromagnetic order of pseudo spins.

As shown in Fig.~\ref{Fig:Illustrator}(c), we create a vortex pair by applying $S^x$ operator on two adjacent sites to simultaneously flip their spin orientations. We note that any closed loop enclosing only the vortex defect [red dot in Fig.~\ref{Fig:Illustrator}(c)] leads to a winding number 1 (modulo $2\pi$), while those around the anti-vortex (black dot) to $-1$. Zero winding number can be counted when a pair of defects (or no defects at all) are enclosed by the loop. Moreover, one can further move the vortex on the triangular lattice, like in a ``tight-binding" model, by flipping spins on further neighboring sites, which naturally leads to a quadratic-type low-energy dispersion near M point along $\Gamma$-M-$\Gamma$ (see Supplementary Note 7). Since the ghost peak in Fig.~\ref{Fig:Sq}(c) only suggests a short-range stripe correlation, the vortex pair can thus move only within a small cluster with incipient stripe order, i.e., they are bounded. The vortex pair only unbinds at the upper KT transition $\Th$ where vortices are proliferated \cite{Kosterlitz1973, Kosterlitz1974}, as seen by the ``softening" of M rotonlike mode in dynamical spectrum in Fig.~\ref{Fig:Dynamics}. 

To conclude, in this work, we have established a protocol of understanding and explaining experiments of frustrated magnets in an unbiased manner, with XTRG and QMC-SAC machinery. The thermodynamic and dynamic results of TMGO are captured to great accuracy, thus allowing comprehensive studies of KT physics therein. At intermediate temperature, the KT phase of TMGO realizes a magnetic analog of 2D superfluid phase with several intriguing properties: (i) There emerges a spin XY symmetry and correspondingly complex order parameter $\psi$, which bears quasi-long-range correlation and phase coherence; (ii) The finite-$T$ spin spectrum contains the long-wave-length magnon and competing gapped rotonlike modes near the BZ boundary with energy signifying the binding of vortex-antivortex pair, which plays a key role in determining finite-$T$ phase of the system; (iii) The quasi-long-range XY order melts and the TMGO becomes paramagnetic as $T$ is above $\Th$, driven by the proliferation of vortex excitations in analogy to the superfluid transition in a helium thin film \cite{Bishop1978,Minnhagen1987}. We note that, different from the liquid helium, TMGO has two-temperature scales that outline the intermediate-temperature KT phase. Similar separation of scales has been seen in other quantum magnetic materials, by temperature \cite{Banerjee2016,Chen2019} or spatial dimensions \cite{Lake2005}, but the KT phase is the first time to be seen. Furthermore, the $\Tl^*, \Th^*$ in the present work constitute tentative estimated of two KT transition temperatures roughly from thermodynamics, which still needs to be precisely determined both numerically and experimentally in the future.

The extraction of anomalous exponents $\eta(T)$ constitutes another  interesting future study. The exponent $\eta(T)$ of the KT phase appears in the algebraic correlations in $\langle S^z_i S^z_j \rangle \sim |r_i-r_j|^{-\eta(T)}$ and seems rather indirect to measure in solid-state experiements. Nevertheless, as discussed above in Fig.~\ref{Fig:Sus}, $\eta(T)$ can be determined from the uniform susceptibility $\chi$, a routine magnetic quantity in experiments. Moreover, it would be interesting to check several distinct predictions of KT physics in this 2D magnetic material. One renowned phenomenon is the universal jump in superfluid density at the KT transition \cite{Nelson1977}, as observed experimentally in helium film \cite{Bishop1978}. Through calculations of the $q$-clock model ($q=5,6$), people have revealed universal jumps in the spin stiffness at both upper and lower KT transitions~\cite{Chatelain2014}, which certainly is an interesting prediction to check in TMGO. Dynamically, nuclear magnetic resonance measurements of relaxation time 1/$T_1$ can be conducted, which probe signals of low-energy magnetic dissipations at the KT transition where vortices proliferate. Besides, non-equilibrium thermodynamics such as thermal transport, would also be very worthwhile to explore in the TMGO magnet.

\section*{Methods}
\textbf{Quantum manybody computations.}
In this work, we combined two many-body numerical approaches: quantum Monte Carlo (QMC)~\cite{YCWang2017} 
and the exponential tensor renormalization group (XTRG), the latter method is 
recently introduced based on matrix product operators (MPOs) 
and logarithmic temperature scales~\cite{Chen2018}. XTRG is employed to simulate the
TLI down to temperatures $T<0.5$ K on YC\,$W\times L$ geometries
up to width $W=9$ with various lengths up to $L=12$.
Both dynamical and equilibrium properties are simulated, with the purpose of
fitting the experimental data and obtaining the right parameters as well as to make some predictions.
The QMC is performed in the space-time lattice of $L \times L \times L_\tau$, where $L=36$
and $L_\tau=\beta/\Delta\tau$ with $\Delta\tau=0.05$ and $\beta \equiv 1/T$.
The space-time configuration is written in the $S^{z}_{i,\tau}$ basis with
both local and Wolff-cluster updates to overcome the long autocorrelation time. 
Since the QMC method is standard, we will only introduce the stochastic analytic continuation (SAC) 
scheme below and leave the QMC itself to the Supplementary  Note 8.

\textbf{Exponential thermal tensor network method.}
For the calculations of equilibrium properties, we start from an high-$T$ density matrix $\hat\rho(\Delta \tau) = e^{- \Delta \tau H}$, whose
MPO representation can be obtained conveniently and accurately (up to machine precision), at a small $\Delta \tau \sim 10^{-3 \sim -4}$. 
One way to obtain such an accurate MPO representation is to exploit the 
series-expansion thermal tensor network technique \cite{Chen.b+:2017:SETTN} via the expansion
\begin{equation}
\hat\rho(\Delta \tau) = e^{- \Delta \tau \hat H} = \sum_{n=0}^{N_c} \frac{(- \Delta \tau)^n}{n!} \hat H^n.
\end{equation}

Given the $\hat \rho(\Delta \tau)$ representation, traditionally one evolves $\hat\rho(\beta)$ 
linear in $\beta$ to reach various lower temperatures, i.e.,
$\beta = L_\tau \Delta \tau$ increases by a small value $\Delta \tau$ 
after each step by multiplying $\hat\rho(\Delta \tau)$ to the density matrix \cite{Li.w+:2011:LTRG}.
However, this linear scheme is not optimal in certain aspects, and encounters challenges in generalization to 2D.
Instead, recent study shows that the block entanglement entropy of MPO is bound by $S_E \leq a \ln \beta + \mathrm{const.}$ 
at a conformal critical point, with $a$ an universal coefficient proportional to the central charge \cite{Chen2018}. 
This suggests an exponential procedure of performing cooling procedure. 
Based on this idea, we have developed the XTRG method,
which turns out to be highly efficient in simulating both 1D critical quantum chains and
various 2D lattice systems \cite{Chen2018,Chen2019,Lih2019}.

In XTRG, we cool down the system by multiplying the thermal state by itself, i.e.,
$\rho_0 \equiv \rho(\Delta \tau)$,  $\rho_1 \equiv \rho_0 \cdot \rho_0 = \rho(2 \Delta\tau)$,
thus $ \rho_n \equiv \rho_{n-1} \cdot \rho_{n-1} =  \rho(2^n \Delta\tau)$,
and reach the low-$T$ thermal states exponentially fast. Efficient compression of MPO bonds is then required to
maintain the cooling procedure, where a truncation scheme optimizing the free energy,
and in the mean time maintaining the thermal entanglement, is involved.
One advantage of XTRG is the convenience and high efficiency to deal 
with long-range interactions after the quasi-1D mapping.
For the TLI model with nearest- ($J_1$) and next-nearest-neighbor ($J_2$) 
interactions considered in this work,
we map the 2D lattice into a quasi-1D geometry following a snake-like path.
The Hamiltonian thus contains ``long-range'' interactions and has an efficient MPO representation
with geometric bond dimension $D_H = 2 W+2$, with $W$ the width of the lattice.
In XTRG calculations, the computational costs scale with power $O(D^4)$,
with $D$ the retained bond dimension in MPO,
which is chosen as large as $500$-$600$  in the present study,
assuring accurate thermodynamical results down to sub-Kelvin regime.

\textbf{QMC-SAC approach.}
We exploit the path integral QMC~\cite{YCWang2017}, equipped with SAC approach, to compute the dynamical properties. The time displaced correlated function, defined as $G(\tau)=\langle S^{z}(\tau) S^{z}(0)\rangle$, for a set of imaginary times $\tau_i \ (i=0,1,\cdots,L_\tau)$ with statistical errors can be obtained from QMC simulations. By SAC method~\cite{Sandvik2016,YQQin2017,CJHuang2018,GYSun2018}, the corresponding real-frequency spectral function $S(\omega)$ can be obtained via $S(\tau)=\int_{-\infty}^\infty d\omega S(\omega)K(\tau,\omega)$, where the kernel $K(\tau,\omega)$ depends on the type of the spectral function, i.e., fermionic or bosonic, finite or zero temperature. The spectra at positive and negative frequencies obey the relation of $S(-\omega)=e^{-\beta\omega}S(\omega)$ and we are restricted at the positive frequencies and the kernel can therefore be written as 
$K(\tau,\omega)=\frac1\pi(e^{-\tau\omega}+e^{-(\beta-\tau)\omega})$.
In order to work with a spectral function that is itself normalized to unity, we further modify the kernel and the spectral function and arrive at the transformation between the imaginary time Green's function $G(\bf{q},\tau)$ and real-frequency spectral function $B(\bf{q},\omega)$
\begin{eqnarray}
\label{eq:eq7}
G(\bf{q},\tau)=\int_0^\infty \frac{d\omega}{\pi}\frac{e^{-\tau\omega}+e^{-(\beta-\tau)\omega}}{1+e^{-\beta\omega}}B(\bf{q},\omega)
\label{Eq:GF}
\end{eqnarray}
where $B(\bf{q},\omega)=S(\bf{q},\omega)(1+e^{-\beta\omega})$.

In the practical calculation, we parametrize the $B(\bf{q}, \omega)$ with a large number of equal-amplitude
$\delta$-functions sampled at locations in a frequency continuum as $B(\omega)=\sum_{i=0}^{N_\omega-1}a_i\delta(\omega-\omega_i)$. 
Then the relationship between Green's function obtained from Eq.~(\ref{Eq:GF}) and from QMC can be described by the goodness of fit $\chi^2$, i.e. $\chi^2=\sum_{i=1}^{N_\tau}\sum_{j=1}^{N_\tau} (G_i-\bar G_i) C_{ij}^{-1}(G_j-\bar G_j)$, 
where $\bar G_i$ is the average of QMC measurement and $C_{ij}$
is covariance matrix $C_{ij}=\frac{1}{N_B(N_B-1)}\sum_{b=1}^{N_B}(G_i^b-\bar G_i)(G_j^b-\bar G_j)$, with $N_B$ the number of bins. Then we update the series of $\delta$-functions in a Metropolis process, from $(a_i,\omega_i)$ to $(a'_i,\omega'_i)$, 
to get a more probable configuration of $B(\bf{q},\omega)$. 
The weight for a given spectrum follows the Boltzmann distribution $P(B)\propto \exp(-\chi^2/2\Theta)$, with $\Theta$ a fictitious
temperature chosen in an optimal way so as to give a statistically sound mean $\chi^2$ value, while still staying in the
regime of significant fluctuations of the sampled spectra
so that a smooth averaged spectral function is obtained. 
The resulting spectra will be collected as an ensemble average of the 
Metropolis process within the configurational space of $\{a_i, \omega_i\}$, 
as detailed in Refs.~\onlinecite{Sandvik2016,YQQin2017,CJHuang2018,GYSun2018}.

\section*{Data availability} 
The data that support the findings of this study are available from the corresponding author upon reasonable request.

\section*{Code availability} 
All numerical codes in this paper are available upon request to the authors. \\

\bibliography{../TMGORef}

\begin{thebibliography}{59}%
\makeatletter
\providecommand \@ifxundefined [1]{%
 \@ifx{#1\undefined}
}%
\providecommand \@ifnum [1]{%
 \ifnum #1\expandafter \@firstoftwo
 \else \expandafter \@secondoftwo
 \fi
}%
\providecommand \@ifx [1]{%
 \ifx #1\expandafter \@firstoftwo
 \else \expandafter \@secondoftwo
 \fi
}%
\providecommand \natexlab [1]{#1}%
\providecommand \enquote  [1]{``#1''}%
\providecommand \bibnamefont  [1]{#1}%
\providecommand \bibfnamefont [1]{#1}%
\providecommand \citenamefont [1]{#1}%
\providecommand \href@noop [0]{\@secondoftwo}%
\providecommand \href [0]{\begingroup \@sanitize@url \@href}%
\providecommand \@href[1]{\@@startlink{#1}\@@href}%
\providecommand \@@href[1]{\endgroup#1\@@endlink}%
\providecommand \@sanitize@url [0]{\catcode `\\12\catcode `\$12\catcode
  `\&12\catcode `\#12\catcode `\^12\catcode `\_12\catcode `\%12\relax}%
\providecommand \@@startlink[1]{}%
\providecommand \@@endlink[0]{}%
\providecommand \url  [0]{\begingroup\@sanitize@url \@url }%
\providecommand \@url [1]{\endgroup\@href {#1}{\urlprefix }}%
\providecommand \urlprefix  [0]{URL }%
\providecommand \Eprint [0]{\href }%
\providecommand \doibase [0]{http://dx.doi.org/}%
\providecommand \selectlanguage [0]{\@gobble}%
\providecommand \bibinfo  [0]{\@secondoftwo}%
\providecommand \bibfield  [0]{\@secondoftwo}%
\providecommand \translation [1]{[#1]}%
\providecommand \BibitemOpen [0]{}%
\providecommand \bibitemStop [0]{}%
\providecommand \bibitemNoStop [0]{.\EOS\space}%
\providecommand \EOS [0]{\spacefactor3000\relax}%
\providecommand \BibitemShut  [1]{\csname bibitem#1\endcsname}%
\let\auto@bib@innerbib\@empty
\bibitem [{\citenamefont {Mermin}\ and\ \citenamefont
  {Wagner}(1966)}]{Mermin-Wagner}%
  \BibitemOpen
  \bibfield  {author} {\bibinfo {author} {\bibfnamefont {N.~D.}\ \bibnamefont
  {Mermin}}\ and\ \bibinfo {author} {\bibfnamefont {H.}~\bibnamefont
  {Wagner}},\ }\bibfield  {title} {\enquote {\bibinfo {title} {Absence of
  ferromagnetism or antiferromagnetism in one- or two-dimensional isotropic
  {{Heisenberg}} models},}\ }\href {\doibase 10.1103/PhysRevLett.17.1133}
  {\bibfield  {journal} {\bibinfo  {journal} {Phys. Rev. Lett.}\ }\textbf
  {\bibinfo {volume} {17}},\ \bibinfo {pages} {1133--1136} (\bibinfo {year}
  {1966})}\BibitemShut {NoStop}%
\bibitem [{\citenamefont {Kosterlitz}\ and\ \citenamefont
  {Thouless}(1973)}]{Kosterlitz1973}%
  \BibitemOpen
  \bibfield  {author} {\bibinfo {author} {\bibfnamefont {J.~M.}\ \bibnamefont
  {Kosterlitz}}\ and\ \bibinfo {author} {\bibfnamefont {D.~J.}\ \bibnamefont
  {Thouless}},\ }\bibfield  {title} {\enquote {\bibinfo {title} {Ordering,
  metastability and phase transitions in two-dimensional systems},}\ }\href
  {\doibase 10.1088/0022-3719/6/7/010} {\bibfield  {journal} {\bibinfo
  {journal} {J. Phys. C: Solid State Phys.}\ }\textbf {\bibinfo {volume} {6}},\
  \bibinfo {pages} {1181--1203} (\bibinfo {year} {1973})}\BibitemShut {NoStop}%
\bibitem [{\citenamefont {Kosterlitz}(1974)}]{Kosterlitz1974}%
  \BibitemOpen
  \bibfield  {author} {\bibinfo {author} {\bibfnamefont {J.~M.}\ \bibnamefont
  {Kosterlitz}},\ }\bibfield  {title} {\enquote {\bibinfo {title} {The critical
  properties of the two-dimensional xy model},}\ }\href {\doibase
  10.1088/0022-3719/7/6/005} {\bibfield  {journal} {\bibinfo  {journal} {J.
  Phys. C: Solid State Phys.}\ }\textbf {\bibinfo {volume} {7}},\ \bibinfo
  {pages} {1046--1060} (\bibinfo {year} {1974})}\BibitemShut {NoStop}%
\bibitem [{\citenamefont {Bishop}\ and\ \citenamefont
  {Reppy}(1978)}]{Bishop1978}%
  \BibitemOpen
  \bibfield  {author} {\bibinfo {author} {\bibfnamefont {D.~J.}\ \bibnamefont
  {Bishop}}\ and\ \bibinfo {author} {\bibfnamefont {J.~D.}\ \bibnamefont
  {Reppy}},\ }\bibfield  {title} {\enquote {\bibinfo {title} {Study of the
  superfluid transition in two-dimensional $^{4}\mathrm{He}$ films},}\ }\href
  {\doibase 10.1103/PhysRevLett.40.1727} {\bibfield  {journal} {\bibinfo
  {journal} {Phys. Rev. Lett.}\ }\textbf {\bibinfo {volume} {40}},\ \bibinfo
  {pages} {1727--1730} (\bibinfo {year} {1978})}\BibitemShut {NoStop}%
\bibitem [{\citenamefont {Hadzibabic}\ \emph {et~al.}(2006)\citenamefont
  {Hadzibabic}, \citenamefont {Kr{\"{u}}ger}, \citenamefont {Cheneau},
  \citenamefont {Battelier},\ and\ \citenamefont {Dalibard}}]{Hadzibabic2006}%
  \BibitemOpen
  \bibfield  {author} {\bibinfo {author} {\bibfnamefont {Z.}~\bibnamefont
  {Hadzibabic}}, \bibinfo {author} {\bibfnamefont {P.}~\bibnamefont
  {Kr{\"{u}}ger}}, \bibinfo {author} {\bibfnamefont {M.}~\bibnamefont
  {Cheneau}}, \bibinfo {author} {\bibfnamefont {B.}~\bibnamefont {Battelier}},
  \ and\ \bibinfo {author} {\bibfnamefont {J.}~\bibnamefont {Dalibard}},\
  }\bibfield  {title} {\enquote {\bibinfo {title}
  {{{Berezinskii}-{Kosterlitz}-{Thouless} crossover in a trapped atomic
  gas}},}\ }\href {\doibase 10.1038/nature04851} {\bibfield  {journal}
  {\bibinfo  {journal} {Nature}\ }\textbf {\bibinfo {volume} {441}},\ \bibinfo
  {pages} {1118--1121} (\bibinfo {year} {2006})}\BibitemShut {NoStop}%
\bibitem [{\citenamefont {Desbuquois}\ \emph {et~al.}(2012)\citenamefont
  {Desbuquois}, \citenamefont {Chomaz}, \citenamefont {Yefsah}, \citenamefont
  {L{\'{e}}onard}, \citenamefont {Beugnon}, \citenamefont {Weitenberg},\ and\
  \citenamefont {Dalibard}}]{Desbuquois2012}%
  \BibitemOpen
  \bibfield  {author} {\bibinfo {author} {\bibfnamefont {R.}~\bibnamefont
  {Desbuquois}}, \bibinfo {author} {\bibfnamefont {L.}~\bibnamefont {Chomaz}},
  \bibinfo {author} {\bibfnamefont {T.}~\bibnamefont {Yefsah}}, \bibinfo
  {author} {\bibfnamefont {J.}~\bibnamefont {L{\'{e}}onard}}, \bibinfo {author}
  {\bibfnamefont {J.}~\bibnamefont {Beugnon}}, \bibinfo {author} {\bibfnamefont
  {C.}~\bibnamefont {Weitenberg}}, \ and\ \bibinfo {author} {\bibfnamefont
  {J.}~\bibnamefont {Dalibard}},\ }\bibfield  {title} {\enquote {\bibinfo
  {title} {{Superfluid behaviour of a two-dimensional Bose gas}},}\ }\href
  {https://doi.org/10.1038/nphys2378 http://10.0.4.14/nphys2378} {\bibfield
  {journal} {\bibinfo  {journal} {Nature Phys.}\ }\textbf {\bibinfo {volume}
  {8}},\ \bibinfo {pages} {645--648} (\bibinfo {year} {2012})}\BibitemShut
  {NoStop}%
\bibitem [{\citenamefont {Landau}(1949)}]{Landau1949}%
  \BibitemOpen
  \bibfield  {author} {\bibinfo {author} {\bibfnamefont {L.}~\bibnamefont
  {Landau}},\ }\bibfield  {title} {\enquote {\bibinfo {title} {On the theory of
  superfluidity},}\ }\href {\doibase 10.1103/PhysRev.75.884} {\bibfield
  {journal} {\bibinfo  {journal} {Phys. Rev.}\ }\textbf {\bibinfo {volume}
  {75}},\ \bibinfo {pages} {884--885} (\bibinfo {year} {1949})}\BibitemShut
  {NoStop}%
\bibitem [{\citenamefont {Feynman}(1955)}]{Feynman1955}%
  \BibitemOpen
  \bibfield  {author} {\bibinfo {author} {\bibfnamefont {R.~P.}\ \bibnamefont
  {Feynman}},\ }\href {https://doi.org/10.1016/S0079-6417(08)60077-3} {\emph
  {\bibinfo {title} {Progress in Low Temperature Physics: Chapter II,
  Application of Quantum Mechanics to Liquid Helium}}},\ edited by\ \bibinfo
  {editor} {\bibfnamefont {C.~G.}\ \bibnamefont {Gorter}},\ Vol.~\bibinfo
  {volume} {1}\ (\bibinfo  {publisher} {North Holland},\ \bibinfo {year}
  {1955})\BibitemShut {NoStop}%
\bibitem [{\citenamefont {Bendt}\ \emph {et~al.}(1959)\citenamefont {Bendt},
  \citenamefont {Cowan},\ and\ \citenamefont {Yarnell}}]{Bendt1959}%
  \BibitemOpen
  \bibfield  {author} {\bibinfo {author} {\bibfnamefont {P.~J.}\ \bibnamefont
  {Bendt}}, \bibinfo {author} {\bibfnamefont {R.~D.}\ \bibnamefont {Cowan}}, \
  and\ \bibinfo {author} {\bibfnamefont {J.~L.}\ \bibnamefont {Yarnell}},\
  }\bibfield  {title} {\enquote {\bibinfo {title} {Excitations in liquid
  {Helium}: Thermodynamic calculations},}\ }\href {\doibase
  10.1103/PhysRev.113.1386} {\bibfield  {journal} {\bibinfo  {journal} {Phys.
  Rev.}\ }\textbf {\bibinfo {volume} {113}},\ \bibinfo {pages} {1386--1395}
  (\bibinfo {year} {1959})}\BibitemShut {NoStop}%
\bibitem [{\citenamefont {Isakov}\ and\ \citenamefont
  {Moessner}(2003)}]{Isakov2003}%
  \BibitemOpen
  \bibfield  {author} {\bibinfo {author} {\bibfnamefont {S.~V.}\ \bibnamefont
  {Isakov}}\ and\ \bibinfo {author} {\bibfnamefont {R.}~\bibnamefont
  {Moessner}},\ }\bibfield  {title} {\enquote {\bibinfo {title} {Interplay of
  quantum and thermal fluctuations in a frustrated magnet},}\ }\href {\doibase
  10.1103/PhysRevB.68.104409} {\bibfield  {journal} {\bibinfo  {journal} {Phys.
  Rev. B}\ }\textbf {\bibinfo {volume} {68}},\ \bibinfo {pages} {104409}
  (\bibinfo {year} {2003})}\BibitemShut {NoStop}%
\bibitem [{\citenamefont {Zhou}\ \emph {et~al.}(2012)\citenamefont {Zhou},
  \citenamefont {Xu}, \citenamefont {Hallas}, \citenamefont {Silverstein},
  \citenamefont {Wiebe}, \citenamefont {Umegaki}, \citenamefont {Yan},
  \citenamefont {Murphy}, \citenamefont {Park}, \citenamefont {Qiu},
  \citenamefont {Copley}, \citenamefont {Gardner},\ and\ \citenamefont
  {Takano}}]{Zhou2012}%
  \BibitemOpen
  \bibfield  {author} {\bibinfo {author} {\bibfnamefont {H.~D.}\ \bibnamefont
  {Zhou}}, \bibinfo {author} {\bibfnamefont {C.}~\bibnamefont {Xu}}, \bibinfo
  {author} {\bibfnamefont {A.~M.}\ \bibnamefont {Hallas}}, \bibinfo {author}
  {\bibfnamefont {H.~J.}\ \bibnamefont {Silverstein}}, \bibinfo {author}
  {\bibfnamefont {C.~R.}\ \bibnamefont {Wiebe}}, \bibinfo {author}
  {\bibfnamefont {I.}~\bibnamefont {Umegaki}}, \bibinfo {author} {\bibfnamefont
  {J.~Q.}\ \bibnamefont {Yan}}, \bibinfo {author} {\bibfnamefont {T.~P.}\
  \bibnamefont {Murphy}}, \bibinfo {author} {\bibfnamefont {J.-H.}\
  \bibnamefont {Park}}, \bibinfo {author} {\bibfnamefont {Y.}~\bibnamefont
  {Qiu}}, \bibinfo {author} {\bibfnamefont {J.~R.~D.}\ \bibnamefont {Copley}},
  \bibinfo {author} {\bibfnamefont {J.~S.}\ \bibnamefont {Gardner}}, \ and\
  \bibinfo {author} {\bibfnamefont {Y.}~\bibnamefont {Takano}},\ }\bibfield
  {title} {\enquote {\bibinfo {title} {Successive phase transitions and
  extended spin-excitation continuum in the ${S}\mathbf{=}\frac{1}{2}$
  triangular-lattice antiferromagnet {Ba$_3$CoSb$_2$O$_9$}},}\ }\href {\doibase
  10.1103/PhysRevLett.109.267206} {\bibfield  {journal} {\bibinfo  {journal}
  {Phys. Rev. Lett.}\ }\textbf {\bibinfo {volume} {109}},\ \bibinfo {pages}
  {267206} (\bibinfo {year} {2012})}\BibitemShut {NoStop}%
\bibitem [{\citenamefont {Susuki}\ \emph {et~al.}(2013)\citenamefont {Susuki},
  \citenamefont {Kurita}, \citenamefont {Tanaka}, \citenamefont {Nojiri},
  \citenamefont {Matsuo}, \citenamefont {Kindo},\ and\ \citenamefont
  {Tanaka}}]{Susuki2013}%
  \BibitemOpen
  \bibfield  {author} {\bibinfo {author} {\bibfnamefont {T.}~\bibnamefont
  {Susuki}}, \bibinfo {author} {\bibfnamefont {N.}~\bibnamefont {Kurita}},
  \bibinfo {author} {\bibfnamefont {T.}~\bibnamefont {Tanaka}}, \bibinfo
  {author} {\bibfnamefont {H.}~\bibnamefont {Nojiri}}, \bibinfo {author}
  {\bibfnamefont {A.}~\bibnamefont {Matsuo}}, \bibinfo {author} {\bibfnamefont
  {K.}~\bibnamefont {Kindo}}, \ and\ \bibinfo {author} {\bibfnamefont
  {H.}~\bibnamefont {Tanaka}},\ }\bibfield  {title} {\enquote {\bibinfo {title}
  {Magnetization process and collective excitations in the {$S\mathbf{=}1/2$}
  triangular-lattice {Heisenberg} antiferromagnet {Ba$_3$CoSb$_2$O$_9$}},}\
  }\href {\doibase 10.1103/PhysRevLett.110.267201} {\bibfield  {journal}
  {\bibinfo  {journal} {Phys. Rev. Lett.}\ }\textbf {\bibinfo {volume} {110}},\
  \bibinfo {pages} {267201} (\bibinfo {year} {2013})}\BibitemShut {NoStop}%
\bibitem [{\citenamefont {Ma}\ \emph {et~al.}(2016)\citenamefont {Ma},
  \citenamefont {Kamiya}, \citenamefont {Hong}, \citenamefont {Cao},
  \citenamefont {Ehlers}, \citenamefont {Tian}, \citenamefont {Batista},
  \citenamefont {Dun}, \citenamefont {Zhou},\ and\ \citenamefont
  {Matsuda}}]{Ma2016}%
  \BibitemOpen
  \bibfield  {author} {\bibinfo {author} {\bibfnamefont {J.}~\bibnamefont
  {Ma}}, \bibinfo {author} {\bibfnamefont {Y.}~\bibnamefont {Kamiya}}, \bibinfo
  {author} {\bibfnamefont {T.}~\bibnamefont {Hong}}, \bibinfo {author}
  {\bibfnamefont {H.~B.}\ \bibnamefont {Cao}}, \bibinfo {author} {\bibfnamefont
  {G.}~\bibnamefont {Ehlers}}, \bibinfo {author} {\bibfnamefont
  {W.}~\bibnamefont {Tian}}, \bibinfo {author} {\bibfnamefont {C.~D.}\
  \bibnamefont {Batista}}, \bibinfo {author} {\bibfnamefont {Z.~L.}\
  \bibnamefont {Dun}}, \bibinfo {author} {\bibfnamefont {H.~D.}\ \bibnamefont
  {Zhou}}, \ and\ \bibinfo {author} {\bibfnamefont {M.}~\bibnamefont
  {Matsuda}},\ }\bibfield  {title} {\enquote {\bibinfo {title} {Static and
  dynamical properties of the spin-$1/2$ equilateral triangular-lattice
  antiferromagnet {Ba$_3$CoSb$_2$O$_9$}},}\ }\href {\doibase
  10.1103/PhysRevLett.116.087201} {\bibfield  {journal} {\bibinfo  {journal}
  {Phys. Rev. Lett.}\ }\textbf {\bibinfo {volume} {116}},\ \bibinfo {pages}
  {087201} (\bibinfo {year} {2016})}\BibitemShut {NoStop}%
\bibitem [{\citenamefont {Ito}\ \emph {et~al.}(2017)\citenamefont {Ito},
  \citenamefont {Kurita}, \citenamefont {Tanaka}, \citenamefont
  {Ohira-Kawamura}, \citenamefont {Nakajima}, \citenamefont {Itoh},
  \citenamefont {Kuwahara},\ and\ \citenamefont {Kakurai}}]{Ito2017}%
  \BibitemOpen
  \bibfield  {author} {\bibinfo {author} {\bibfnamefont {S.}~\bibnamefont
  {Ito}}, \bibinfo {author} {\bibfnamefont {N.}~\bibnamefont {Kurita}},
  \bibinfo {author} {\bibfnamefont {H.}~\bibnamefont {Tanaka}}, \bibinfo
  {author} {\bibfnamefont {S.}~\bibnamefont {Ohira-Kawamura}}, \bibinfo
  {author} {\bibfnamefont {K.}~\bibnamefont {Nakajima}}, \bibinfo {author}
  {\bibfnamefont {S.}~\bibnamefont {Itoh}}, \bibinfo {author} {\bibfnamefont
  {K.}~\bibnamefont {Kuwahara}}, \ and\ \bibinfo {author} {\bibfnamefont
  {K.}~\bibnamefont {Kakurai}},\ }\bibfield  {title} {\enquote {\bibinfo
  {title} {Structure of the magnetic excitations in the spin-1/2
  triangular-lattice {{Heisenberg}} antiferromagnet {Ba$_3$CoSb$_2$O$_9$}},}\
  }\href {https://www.nature.com/articles/s41467-017-00316-x} {\bibfield
  {journal} {\bibinfo  {journal} {Nat. Commun.}\ }\textbf {\bibinfo {volume}
  {8}},\ \bibinfo {pages} {235} (\bibinfo {year} {2017})}\BibitemShut {NoStop}%
\bibitem [{\citenamefont {Rawl}\ \emph {et~al.}(2017)\citenamefont {Rawl},
  \citenamefont {Ge}, \citenamefont {Agrawal}, \citenamefont {Kamiya},
  \citenamefont {Dela~Cruz}, \citenamefont {Butch}, \citenamefont {Sun},
  \citenamefont {Lee}, \citenamefont {Choi}, \citenamefont {Oitmaa},
  \citenamefont {Batista}, \citenamefont {Mourigal}, \citenamefont {Zhou},\
  and\ \citenamefont {Ma}}]{Rawl2017}%
  \BibitemOpen
  \bibfield  {author} {\bibinfo {author} {\bibfnamefont {R.}~\bibnamefont
  {Rawl}}, \bibinfo {author} {\bibfnamefont {L.}~\bibnamefont {Ge}}, \bibinfo
  {author} {\bibfnamefont {H.}~\bibnamefont {Agrawal}}, \bibinfo {author}
  {\bibfnamefont {Y.}~\bibnamefont {Kamiya}}, \bibinfo {author} {\bibfnamefont
  {C.~R.}\ \bibnamefont {Dela~Cruz}}, \bibinfo {author} {\bibfnamefont {N.~P.}\
  \bibnamefont {Butch}}, \bibinfo {author} {\bibfnamefont {X.~F.}\ \bibnamefont
  {Sun}}, \bibinfo {author} {\bibfnamefont {M.}~\bibnamefont {Lee}}, \bibinfo
  {author} {\bibfnamefont {E.~S.}\ \bibnamefont {Choi}}, \bibinfo {author}
  {\bibfnamefont {J.}~\bibnamefont {Oitmaa}}, \bibinfo {author} {\bibfnamefont
  {C.~D.}\ \bibnamefont {Batista}}, \bibinfo {author} {\bibfnamefont
  {M.}~\bibnamefont {Mourigal}}, \bibinfo {author} {\bibfnamefont {H.~D.}\
  \bibnamefont {Zhou}}, \ and\ \bibinfo {author} {\bibfnamefont
  {J.}~\bibnamefont {Ma}},\ }\bibfield  {title} {\enquote {\bibinfo {title}
  {{${\mathrm{Ba}}_{8}{\mathrm{CoNb}}_{6}{\mathrm{O}}_{24}$}: A
  spin-$\frac{1}{2}$ triangular-lattice {Heisenberg} antiferromagnet in the
  two-dimensional limit},}\ }\href {\doibase 10.1103/PhysRevB.95.060412}
  {\bibfield  {journal} {\bibinfo  {journal} {Phys. Rev. B}\ }\textbf {\bibinfo
  {volume} {95}},\ \bibinfo {pages} {060412(R)} (\bibinfo {year}
  {2017})}\BibitemShut {NoStop}%
\bibitem [{\citenamefont {Cui}\ \emph {et~al.}(2018)\citenamefont {Cui},
  \citenamefont {Dai}, \citenamefont {Zhou}, \citenamefont {Wang},
  \citenamefont {Li}, \citenamefont {Song}, \citenamefont {Wang}, \citenamefont
  {Ma}, \citenamefont {Zhang}, \citenamefont {Li}, \citenamefont {Luke},
  \citenamefont {Normand}, \citenamefont {Xiang},\ and\ \citenamefont
  {Yu}}]{Cui2018}%
  \BibitemOpen
  \bibfield  {author} {\bibinfo {author} {\bibfnamefont {Y.}~\bibnamefont
  {Cui}}, \bibinfo {author} {\bibfnamefont {J.}~\bibnamefont {Dai}}, \bibinfo
  {author} {\bibfnamefont {P.}~\bibnamefont {Zhou}}, \bibinfo {author}
  {\bibfnamefont {P.~S.}\ \bibnamefont {Wang}}, \bibinfo {author}
  {\bibfnamefont {T.~R.}\ \bibnamefont {Li}}, \bibinfo {author} {\bibfnamefont
  {W.~H.}\ \bibnamefont {Song}}, \bibinfo {author} {\bibfnamefont {J.~C.}\
  \bibnamefont {Wang}}, \bibinfo {author} {\bibfnamefont {L.}~\bibnamefont
  {Ma}}, \bibinfo {author} {\bibfnamefont {Z.}~\bibnamefont {Zhang}}, \bibinfo
  {author} {\bibfnamefont {S.~Y.}\ \bibnamefont {Li}}, \bibinfo {author}
  {\bibfnamefont {G.~M.}\ \bibnamefont {Luke}}, \bibinfo {author}
  {\bibfnamefont {B.}~\bibnamefont {Normand}}, \bibinfo {author} {\bibfnamefont
  {T.}~\bibnamefont {Xiang}}, \ and\ \bibinfo {author} {\bibfnamefont
  {W.}~\bibnamefont {Yu}},\ }\bibfield  {title} {\enquote {\bibinfo {title}
  {Mermin-{Wagner} physics, {$(H,T)$} phase diagram, and candidate quantum
  spin-liquid phase in the spin-$\frac{1}{2}$ triangular-lattice
  antiferromagnet {${\mathrm{Ba}}_{8}{\mathrm{CoNb}}_{6}{\mathrm{O}}_{24}$}},}\
  }\href {\doibase 10.1103/PhysRevMaterials.2.044403} {\bibfield  {journal}
  {\bibinfo  {journal} {Phys. Rev. Materials}\ }\textbf {\bibinfo {volume}
  {2}},\ \bibinfo {pages} {044403} (\bibinfo {year} {2018})}\BibitemShut
  {NoStop}%
\bibitem [{\citenamefont {Li}\ \emph {et~al.}(2015)\citenamefont {Li},
  \citenamefont {Chen}, \citenamefont {Tong}, \citenamefont {Pi}, \citenamefont
  {Liu}, \citenamefont {Yang}, \citenamefont {Wang},\ and\ \citenamefont
  {Zhang}}]{Li2015b}%
  \BibitemOpen
  \bibfield  {author} {\bibinfo {author} {\bibfnamefont {Y.}~\bibnamefont
  {Li}}, \bibinfo {author} {\bibfnamefont {G.}~\bibnamefont {Chen}}, \bibinfo
  {author} {\bibfnamefont {W.}~\bibnamefont {Tong}}, \bibinfo {author}
  {\bibfnamefont {L.}~\bibnamefont {Pi}}, \bibinfo {author} {\bibfnamefont
  {J.}~\bibnamefont {Liu}}, \bibinfo {author} {\bibfnamefont {Z.}~\bibnamefont
  {Yang}}, \bibinfo {author} {\bibfnamefont {X.}~\bibnamefont {Wang}}, \ and\
  \bibinfo {author} {\bibfnamefont {Q.}~\bibnamefont {Zhang}},\ }\bibfield
  {title} {\enquote {\bibinfo {title} {Rare-earth triangular lattice spin
  liquid: A single-crystal study of {YbMgGaO$_4$}},}\ }\href {\doibase
  10.1103/PhysRevLett.115.167203} {\bibfield  {journal} {\bibinfo  {journal}
  {Phys. Rev. Lett.}\ }\textbf {\bibinfo {volume} {115}},\ \bibinfo {pages}
  {167203} (\bibinfo {year} {2015})}\BibitemShut {NoStop}%
\bibitem [{\citenamefont {Li}\ \emph {et~al.}(2016)\citenamefont {Li},
  \citenamefont {Adroja}, \citenamefont {Biswas}, \citenamefont {Baker},
  \citenamefont {Zhang}, \citenamefont {Liu}, \citenamefont {Tsirlin},
  \citenamefont {Gegenwart},\ and\ \citenamefont {Zhang}}]{Li2016}%
  \BibitemOpen
  \bibfield  {author} {\bibinfo {author} {\bibfnamefont {Y.}~\bibnamefont
  {Li}}, \bibinfo {author} {\bibfnamefont {D.}~\bibnamefont {Adroja}}, \bibinfo
  {author} {\bibfnamefont {P.~K.}\ \bibnamefont {Biswas}}, \bibinfo {author}
  {\bibfnamefont {P.~J.}\ \bibnamefont {Baker}}, \bibinfo {author}
  {\bibfnamefont {Q.}~\bibnamefont {Zhang}}, \bibinfo {author} {\bibfnamefont
  {J.}~\bibnamefont {Liu}}, \bibinfo {author} {\bibfnamefont {A.~A.}\
  \bibnamefont {Tsirlin}}, \bibinfo {author} {\bibfnamefont {P.}~\bibnamefont
  {Gegenwart}}, \ and\ \bibinfo {author} {\bibfnamefont {Q.}~\bibnamefont
  {Zhang}},\ }\bibfield  {title} {\enquote {\bibinfo {title} {Muon spin
  relaxation evidence for the {U(1)} quantum spin-liquid ground state in the
  triangular antiferromagnet {YbMgGaO$_4$}},}\ }\href {\doibase
  10.1103/PhysRevLett.117.097201} {\bibfield  {journal} {\bibinfo  {journal}
  {Phys. Rev. Lett.}\ }\textbf {\bibinfo {volume} {117}},\ \bibinfo {pages}
  {097201} (\bibinfo {year} {2016})}\BibitemShut {NoStop}%
\bibitem [{\citenamefont {Shen}\ \emph {et~al.}(2016)\citenamefont {Shen},
  \citenamefont {Li}, \citenamefont {Wo}, \citenamefont {Li}, \citenamefont
  {Shen}, \citenamefont {Pan}, \citenamefont {Wang}, \citenamefont {Walker},
  \citenamefont {Steffens}, \citenamefont {Boehm}, \citenamefont {Hao},
  \citenamefont {Quintero-Castro}, \citenamefont {Harriger}, \citenamefont
  {Frontzek}, \citenamefont {Hao}, \citenamefont {Meng}, \citenamefont {Zhang},
  \citenamefont {Chen},\ and\ \citenamefont {Zhao}}]{Shen2016}%
  \BibitemOpen
  \bibfield  {author} {\bibinfo {author} {\bibfnamefont {Y.}~\bibnamefont
  {Shen}}, \bibinfo {author} {\bibfnamefont {Y.-D.}\ \bibnamefont {Li}},
  \bibinfo {author} {\bibfnamefont {H.}~\bibnamefont {Wo}}, \bibinfo {author}
  {\bibfnamefont {Y.}~\bibnamefont {Li}}, \bibinfo {author} {\bibfnamefont
  {S.}~\bibnamefont {Shen}}, \bibinfo {author} {\bibfnamefont {B.}~\bibnamefont
  {Pan}}, \bibinfo {author} {\bibfnamefont {Q.}~\bibnamefont {Wang}}, \bibinfo
  {author} {\bibfnamefont {H.~C.}\ \bibnamefont {Walker}}, \bibinfo {author}
  {\bibfnamefont {P.}~\bibnamefont {Steffens}}, \bibinfo {author}
  {\bibfnamefont {M.}~\bibnamefont {Boehm}}, \bibinfo {author} {\bibfnamefont
  {Y.}~\bibnamefont {Hao}}, \bibinfo {author} {\bibfnamefont {D.~L.}\
  \bibnamefont {Quintero-Castro}}, \bibinfo {author} {\bibfnamefont {L.~W.}\
  \bibnamefont {Harriger}}, \bibinfo {author} {\bibfnamefont {M.~D.}\
  \bibnamefont {Frontzek}}, \bibinfo {author} {\bibfnamefont {L.}~\bibnamefont
  {Hao}}, \bibinfo {author} {\bibfnamefont {S.}~\bibnamefont {Meng}}, \bibinfo
  {author} {\bibfnamefont {Q.}~\bibnamefont {Zhang}}, \bibinfo {author}
  {\bibfnamefont {G.}~\bibnamefont {Chen}}, \ and\ \bibinfo {author}
  {\bibfnamefont {J.}~\bibnamefont {Zhao}},\ }\bibfield  {title} {\enquote
  {\bibinfo {title} {Evidence for a spinon {Fermi} surface in a
  triangular-lattice quantum-spin-liquid candidate},}\ }\href
  {https://doi.org/10.1038/nature20614} {\bibfield  {journal} {\bibinfo
  {journal} {Nature}\ }\textbf {\bibinfo {volume} {540}},\ \bibinfo {pages}
  {559--562} (\bibinfo {year} {2016})}\BibitemShut {NoStop}%
\bibitem [{\citenamefont {Li}\ \emph {et~al.}(2019{\natexlab{a}})\citenamefont
  {Li}, \citenamefont {Bachus}, \citenamefont {Liu}, \citenamefont
  {Radelytskyi}, \citenamefont {Bertin}, \citenamefont {Schneidewind},
  \citenamefont {Tokiwa}, \citenamefont {Tsirlin},\ and\ \citenamefont
  {Gegenwart}}]{Li2019}%
  \BibitemOpen
  \bibfield  {author} {\bibinfo {author} {\bibfnamefont {Y.}~\bibnamefont
  {Li}}, \bibinfo {author} {\bibfnamefont {S.}~\bibnamefont {Bachus}}, \bibinfo
  {author} {\bibfnamefont {B.}~\bibnamefont {Liu}}, \bibinfo {author}
  {\bibfnamefont {I.}~\bibnamefont {Radelytskyi}}, \bibinfo {author}
  {\bibfnamefont {A.}~\bibnamefont {Bertin}}, \bibinfo {author} {\bibfnamefont
  {A.}~\bibnamefont {Schneidewind}}, \bibinfo {author} {\bibfnamefont
  {Y.}~\bibnamefont {Tokiwa}}, \bibinfo {author} {\bibfnamefont {A.~A.}\
  \bibnamefont {Tsirlin}}, \ and\ \bibinfo {author} {\bibfnamefont
  {P.}~\bibnamefont {Gegenwart}},\ }\bibfield  {title} {\enquote {\bibinfo
  {title} {Rearrangement of uncorrelated valence bonds evidenced by low-energy
  spin excitations in {YbMgGaO$_4$}},}\ }\href {\doibase
  10.1103/PhysRevLett.122.137201} {\bibfield  {journal} {\bibinfo  {journal}
  {Phys. Rev. Lett.}\ }\textbf {\bibinfo {volume} {122}},\ \bibinfo {pages}
  {137201} (\bibinfo {year} {2019}{\natexlab{a}})}\BibitemShut {NoStop}%
\bibitem [{\citenamefont {Zhu}\ \emph {et~al.}(2017)\citenamefont {Zhu},
  \citenamefont {Maksimov}, \citenamefont {White},\ and\ \citenamefont
  {Chernyshev}}]{Zhu2017}%
  \BibitemOpen
  \bibfield  {author} {\bibinfo {author} {\bibfnamefont {Z.}~\bibnamefont
  {Zhu}}, \bibinfo {author} {\bibfnamefont {P.~A.}\ \bibnamefont {Maksimov}},
  \bibinfo {author} {\bibfnamefont {S.~R.}\ \bibnamefont {White}}, \ and\
  \bibinfo {author} {\bibfnamefont {A.~L.}\ \bibnamefont {Chernyshev}},\
  }\bibfield  {title} {\enquote {\bibinfo {title} {Disorder-induced mimicry of
  a spin liquid in {YbMgGaO$_4$}},}\ }\href {\doibase
  10.1103/PhysRevLett.119.157201} {\bibfield  {journal} {\bibinfo  {journal}
  {Phys. Rev. Lett.}\ }\textbf {\bibinfo {volume} {119}},\ \bibinfo {pages}
  {157201} (\bibinfo {year} {2017})}\BibitemShut {NoStop}%
\bibitem [{\citenamefont {Kimchi}\ \emph {et~al.}(2018)\citenamefont {Kimchi},
  \citenamefont {Nahum},\ and\ \citenamefont {Senthil}}]{Kimchi2018}%
  \BibitemOpen
  \bibfield  {author} {\bibinfo {author} {\bibfnamefont {I.}~\bibnamefont
  {Kimchi}}, \bibinfo {author} {\bibfnamefont {A.}~\bibnamefont {Nahum}}, \
  and\ \bibinfo {author} {\bibfnamefont {T.}~\bibnamefont {Senthil}},\
  }\bibfield  {title} {\enquote {\bibinfo {title} {Valence bonds in random
  quantum magnets: Theory and application to {YbMgGaO$_4$}},}\ }\href {\doibase
  10.1103/PhysRevX.8.031028} {\bibfield  {journal} {\bibinfo  {journal} {Phys.
  Rev. X}\ }\textbf {\bibinfo {volume} {8}},\ \bibinfo {pages} {031028}
  (\bibinfo {year} {2018})}\BibitemShut {NoStop}%
\bibitem [{\citenamefont {Ma}\ \emph {et~al.}(2018)\citenamefont {Ma},
  \citenamefont {Wang}, \citenamefont {Dong}, \citenamefont {Zhang},
  \citenamefont {Li}, \citenamefont {Zheng}, \citenamefont {Yu}, \citenamefont
  {Wang}, \citenamefont {Che}, \citenamefont {Ran}, \citenamefont {Bao},
  \citenamefont {Cai}, \citenamefont {\ifmmode~\check{C}\else
  \v{C}\fi{}erm\'ak}, \citenamefont {Schneidewind}, \citenamefont {Yano},
  \citenamefont {Gardner}, \citenamefont {Lu}, \citenamefont {Yu},
  \citenamefont {Liu}, \citenamefont {Li}, \citenamefont {Li},\ and\
  \citenamefont {Wen}}]{JSWen2018}%
  \BibitemOpen
  \bibfield  {author} {\bibinfo {author} {\bibfnamefont {Z.}~\bibnamefont
  {Ma}}, \bibinfo {author} {\bibfnamefont {J.}~\bibnamefont {Wang}}, \bibinfo
  {author} {\bibfnamefont {Z.-Y.}\ \bibnamefont {Dong}}, \bibinfo {author}
  {\bibfnamefont {J.}~\bibnamefont {Zhang}}, \bibinfo {author} {\bibfnamefont
  {S.}~\bibnamefont {Li}}, \bibinfo {author} {\bibfnamefont {S.-H.}\
  \bibnamefont {Zheng}}, \bibinfo {author} {\bibfnamefont {Y.}~\bibnamefont
  {Yu}}, \bibinfo {author} {\bibfnamefont {W.}~\bibnamefont {Wang}}, \bibinfo
  {author} {\bibfnamefont {L.}~\bibnamefont {Che}}, \bibinfo {author}
  {\bibfnamefont {K.}~\bibnamefont {Ran}}, \bibinfo {author} {\bibfnamefont
  {S.}~\bibnamefont {Bao}}, \bibinfo {author} {\bibfnamefont {Z.}~\bibnamefont
  {Cai}}, \bibinfo {author} {\bibfnamefont {P.}~\bibnamefont
  {\ifmmode~\check{C}\else \v{C}\fi{}erm\'ak}}, \bibinfo {author}
  {\bibfnamefont {A.}~\bibnamefont {Schneidewind}}, \bibinfo {author}
  {\bibfnamefont {S.}~\bibnamefont {Yano}}, \bibinfo {author} {\bibfnamefont
  {J.~S.}\ \bibnamefont {Gardner}}, \bibinfo {author} {\bibfnamefont
  {X.}~\bibnamefont {Lu}}, \bibinfo {author} {\bibfnamefont {S.-L.}\
  \bibnamefont {Yu}}, \bibinfo {author} {\bibfnamefont {J.-M.}\ \bibnamefont
  {Liu}}, \bibinfo {author} {\bibfnamefont {S.}~\bibnamefont {Li}}, \bibinfo
  {author} {\bibfnamefont {J.-X.}\ \bibnamefont {Li}}, \ and\ \bibinfo {author}
  {\bibfnamefont {J.}~\bibnamefont {Wen}},\ }\bibfield  {title} {\enquote
  {\bibinfo {title} {Spin-glass ground state in a triangular-lattice compound
  {YbZnGaO$_{4}$}},}\ }\href {\doibase 10.1103/PhysRevLett.120.087201}
  {\bibfield  {journal} {\bibinfo  {journal} {Phys. Rev. Lett.}\ }\textbf
  {\bibinfo {volume} {120}},\ \bibinfo {pages} {087201} (\bibinfo {year}
  {2018})}\BibitemShut {NoStop}%
\bibitem [{\citenamefont {Cevallos}\ \emph {et~al.}(2018)\citenamefont
  {Cevallos}, \citenamefont {Stolze}, \citenamefont {Kong},\ and\ \citenamefont
  {Cava}}]{Cava2018}%
  \BibitemOpen
  \bibfield  {author} {\bibinfo {author} {\bibfnamefont {F.~A.}\ \bibnamefont
  {Cevallos}}, \bibinfo {author} {\bibfnamefont {K.}~\bibnamefont {Stolze}},
  \bibinfo {author} {\bibfnamefont {T.}~\bibnamefont {Kong}}, \ and\ \bibinfo
  {author} {\bibfnamefont {R.~J.}\ \bibnamefont {Cava}},\ }\bibfield  {title}
  {\enquote {\bibinfo {title} {Anisotropic magnetic properties of the
  triangular plane lattice material {TmMgGaO}$_4$},}\ }\href {\doibase
  https://doi.org/10.1016/j.materresbull.2018.04.042} {\bibfield  {journal}
  {\bibinfo  {journal} {Mater. Res. Bull.}\ }\textbf {\bibinfo {volume}
  {105}},\ \bibinfo {pages} {154--158} (\bibinfo {year} {2018})}\BibitemShut
  {NoStop}%
\bibitem [{\citenamefont {Li}\ \emph {et~al.}(2020)\citenamefont {Li},
  \citenamefont {Bachus}, \citenamefont {Deng}, \citenamefont {Schmidt},
  \citenamefont {Thoma}, \citenamefont {Hutanu}, \citenamefont {Tokiwa},
  \citenamefont {Tsirlin},\ and\ \citenamefont {Gegenwart}}]{Li2018}%
  \BibitemOpen
  \bibfield  {author} {\bibinfo {author} {\bibfnamefont {Y.}~\bibnamefont
  {Li}}, \bibinfo {author} {\bibfnamefont {S.}~\bibnamefont {Bachus}}, \bibinfo
  {author} {\bibfnamefont {H.}~\bibnamefont {Deng}}, \bibinfo {author}
  {\bibfnamefont {W.}~\bibnamefont {Schmidt}}, \bibinfo {author} {\bibfnamefont
  {H.}~\bibnamefont {Thoma}}, \bibinfo {author} {\bibfnamefont
  {V.}~\bibnamefont {Hutanu}}, \bibinfo {author} {\bibfnamefont
  {Y.}~\bibnamefont {Tokiwa}}, \bibinfo {author} {\bibfnamefont {A.~A.}\
  \bibnamefont {Tsirlin}}, \ and\ \bibinfo {author} {\bibfnamefont
  {P.}~\bibnamefont {Gegenwart}},\ }\bibfield  {title} {\enquote {\bibinfo
  {title} {Partial up-up-down order with the continuously distributed order
  parameter in the triangular antiferromagnet {${\mathrm{TmMgGaO}}_{4}$}},}\
  }\href {\doibase 10.1103/PhysRevX.10.011007} {\bibfield  {journal} {\bibinfo
  {journal} {Phys. Rev. X}\ }\textbf {\bibinfo {volume} {10}},\ \bibinfo
  {pages} {011007} (\bibinfo {year} {2020})}\BibitemShut {NoStop}%
\bibitem [{\citenamefont {Shen}\ \emph {et~al.}(2019)\citenamefont {Shen},
  \citenamefont {Liu}, \citenamefont {Qin}, \citenamefont {Shen}, \citenamefont
  {Li}, \citenamefont {Bewley}, \citenamefont {Schneidewind}, \citenamefont
  {Chen},\ and\ \citenamefont {Zhao}}]{Shen2018}%
  \BibitemOpen
  \bibfield  {author} {\bibinfo {author} {\bibfnamefont {Y.}~\bibnamefont
  {Shen}}, \bibinfo {author} {\bibfnamefont {C.}~\bibnamefont {Liu}}, \bibinfo
  {author} {\bibfnamefont {Y.}~\bibnamefont {Qin}}, \bibinfo {author}
  {\bibfnamefont {S.}~\bibnamefont {Shen}}, \bibinfo {author} {\bibfnamefont
  {Y.-D.}\ \bibnamefont {Li}}, \bibinfo {author} {\bibfnamefont
  {R.}~\bibnamefont {Bewley}}, \bibinfo {author} {\bibfnamefont
  {A.}~\bibnamefont {Schneidewind}}, \bibinfo {author} {\bibfnamefont
  {G.}~\bibnamefont {Chen}}, \ and\ \bibinfo {author} {\bibfnamefont
  {J.}~\bibnamefont {Zhao}},\ }\bibfield  {title} {\enquote {\bibinfo {title}
  {Intertwined dipolar and multipolar order in the triangular-lattice magnet
  {TmMgGaO4}},}\ }\href {\doibase 10.1038/s41467-019-12410-3} {\bibfield
  {journal} {\bibinfo  {journal} {Nat. Commun.}\ }\textbf {\bibinfo {volume}
  {10}},\ \bibinfo {pages} {4530} (\bibinfo {year} {2019})}\BibitemShut
  {NoStop}%
\bibitem [{\citenamefont {Chen}\ \emph {et~al.}(2018)\citenamefont {Chen},
  \citenamefont {Chen}, \citenamefont {Chen}, \citenamefont {Li},\ and\
  \citenamefont {Weichselbaum}}]{Chen2018}%
  \BibitemOpen
  \bibfield  {author} {\bibinfo {author} {\bibfnamefont {B.-B.}\ \bibnamefont
  {Chen}}, \bibinfo {author} {\bibfnamefont {L.}~\bibnamefont {Chen}}, \bibinfo
  {author} {\bibfnamefont {Z.}~\bibnamefont {Chen}}, \bibinfo {author}
  {\bibfnamefont {W.}~\bibnamefont {Li}}, \ and\ \bibinfo {author}
  {\bibfnamefont {A.}~\bibnamefont {Weichselbaum}},\ }\bibfield  {title}
  {\enquote {\bibinfo {title} {Exponential thermal tensor network approach for
  quantum lattice models},}\ }\href {\doibase 10.1103/PhysRevX.8.031082}
  {\bibfield  {journal} {\bibinfo  {journal} {Phys. Rev. X}\ }\textbf {\bibinfo
  {volume} {8}},\ \bibinfo {pages} {031082} (\bibinfo {year}
  {2018})}\BibitemShut {NoStop}%
\bibitem [{\citenamefont {Sandvik}(2016)}]{Sandvik2016}%
  \BibitemOpen
  \bibfield  {author} {\bibinfo {author} {\bibfnamefont {A.~W.}\ \bibnamefont
  {Sandvik}},\ }\bibfield  {title} {\enquote {\bibinfo {title} {Constrained
  sampling method for analytic continuation},}\ }\href {\doibase
  10.1103/PhysRevE.94.063308} {\bibfield  {journal} {\bibinfo  {journal} {Phys.
  Rev. E}\ }\textbf {\bibinfo {volume} {94}},\ \bibinfo {pages} {063308}
  (\bibinfo {year} {2016})}\BibitemShut {NoStop}%
\bibitem [{\citenamefont {Qin}\ \emph {et~al.}(2017)\citenamefont {Qin},
  \citenamefont {Normand}, \citenamefont {Sandvik},\ and\ \citenamefont
  {Meng}}]{YQQin2017}%
  \BibitemOpen
  \bibfield  {author} {\bibinfo {author} {\bibfnamefont {Y.~Q.}\ \bibnamefont
  {Qin}}, \bibinfo {author} {\bibfnamefont {B.}~\bibnamefont {Normand}},
  \bibinfo {author} {\bibfnamefont {A.~W.}\ \bibnamefont {Sandvik}}, \ and\
  \bibinfo {author} {\bibfnamefont {Z.~Y.}\ \bibnamefont {Meng}},\ }\bibfield
  {title} {\enquote {\bibinfo {title} {Amplitude mode in three-dimensional
  dimerized antiferromagnets},}\ }\href {\doibase
  10.1103/PhysRevLett.118.147207} {\bibfield  {journal} {\bibinfo  {journal}
  {Phys. Rev. Lett.}\ }\textbf {\bibinfo {volume} {118}},\ \bibinfo {pages}
  {147207} (\bibinfo {year} {2017})}\BibitemShut {NoStop}%
\bibitem [{\citenamefont {Shao}\ \emph {et~al.}(2017)\citenamefont {Shao},
  \citenamefont {Qin}, \citenamefont {Capponi}, \citenamefont {Chesi},
  \citenamefont {Meng},\ and\ \citenamefont {Sandvik}}]{Shao2017}%
  \BibitemOpen
  \bibfield  {author} {\bibinfo {author} {\bibfnamefont {H.}~\bibnamefont
  {Shao}}, \bibinfo {author} {\bibfnamefont {Y.~Q.}\ \bibnamefont {Qin}},
  \bibinfo {author} {\bibfnamefont {S.}~\bibnamefont {Capponi}}, \bibinfo
  {author} {\bibfnamefont {S.}~\bibnamefont {Chesi}}, \bibinfo {author}
  {\bibfnamefont {Z.~Y.}\ \bibnamefont {Meng}}, \ and\ \bibinfo {author}
  {\bibfnamefont {A.~W.}\ \bibnamefont {Sandvik}},\ }\bibfield  {title}
  {\enquote {\bibinfo {title} {Nearly deconfined spinon excitations in the
  square-lattice spin-$1/2$ {Heisenberg} antiferromagnet},}\ }\href {\doibase
  10.1103/PhysRevX.7.041072} {\bibfield  {journal} {\bibinfo  {journal} {Phys.
  Rev. X}\ }\textbf {\bibinfo {volume} {7}},\ \bibinfo {pages} {041072}
  (\bibinfo {year} {2017})}\BibitemShut {NoStop}%
\bibitem [{\citenamefont {Huang}\ \emph {et~al.}(2018)\citenamefont {Huang},
  \citenamefont {Deng}, \citenamefont {Wan},\ and\ \citenamefont
  {Meng}}]{CJHuang2018}%
  \BibitemOpen
  \bibfield  {author} {\bibinfo {author} {\bibfnamefont {C.-J.}\ \bibnamefont
  {Huang}}, \bibinfo {author} {\bibfnamefont {Y.}~\bibnamefont {Deng}},
  \bibinfo {author} {\bibfnamefont {Y.}~\bibnamefont {Wan}}, \ and\ \bibinfo
  {author} {\bibfnamefont {Z.~Y.}\ \bibnamefont {Meng}},\ }\bibfield  {title}
  {\enquote {\bibinfo {title} {Dynamics of topological excitations in a model
  quantum spin ice},}\ }\href {\doibase 10.1103/PhysRevLett.120.167202}
  {\bibfield  {journal} {\bibinfo  {journal} {Phys. Rev. Lett.}\ }\textbf
  {\bibinfo {volume} {120}},\ \bibinfo {pages} {167202} (\bibinfo {year}
  {2018})}\BibitemShut {NoStop}%
\bibitem [{\citenamefont {Sun}\ \emph {et~al.}(2018)\citenamefont {Sun},
  \citenamefont {Wang}, \citenamefont {Fang}, \citenamefont {Qi}, \citenamefont
  {Cheng},\ and\ \citenamefont {Meng}}]{GYSun2018}%
  \BibitemOpen
  \bibfield  {author} {\bibinfo {author} {\bibfnamefont {G.-Y.}\ \bibnamefont
  {Sun}}, \bibinfo {author} {\bibfnamefont {Y.-C.}\ \bibnamefont {Wang}},
  \bibinfo {author} {\bibfnamefont {C.}~\bibnamefont {Fang}}, \bibinfo {author}
  {\bibfnamefont {Y.}~\bibnamefont {Qi}}, \bibinfo {author} {\bibfnamefont
  {M.}~\bibnamefont {Cheng}}, \ and\ \bibinfo {author} {\bibfnamefont {Z.~Y.}\
  \bibnamefont {Meng}},\ }\bibfield  {title} {\enquote {\bibinfo {title}
  {Dynamical signature of symmetry fractionalization in frustrated magnets},}\
  }\href {\doibase 10.1103/PhysRevLett.121.077201} {\bibfield  {journal}
  {\bibinfo  {journal} {Phys. Rev. Lett.}\ }\textbf {\bibinfo {volume} {121}},\
  \bibinfo {pages} {077201} (\bibinfo {year} {2018})}\BibitemShut {NoStop}%
\bibitem [{\citenamefont {Kresse}\ and\ \citenamefont
  {Furthm\"uller}(1996)}]{VASP1}%
  \BibitemOpen
  \bibfield  {author} {\bibinfo {author} {\bibfnamefont {G.}~\bibnamefont
  {Kresse}}\ and\ \bibinfo {author} {\bibfnamefont {J.}~\bibnamefont
  {Furthm\"uller}},\ }\bibfield  {title} {\enquote {\bibinfo {title} {Efficient
  iterative schemes for ab initio total-energy calculations using a plane-wave
  basis set},}\ }\href {\doibase 10.1103/PhysRevB.54.11169} {\bibfield
  {journal} {\bibinfo  {journal} {Phys. Rev. B}\ }\textbf {\bibinfo {volume}
  {54}},\ \bibinfo {pages} {11169--11186} (\bibinfo {year} {1996})}\BibitemShut
  {NoStop}%
\bibitem [{\citenamefont {Kresse}\ and\ \citenamefont {Joubert}(1999)}]{VASP2}%
  \BibitemOpen
  \bibfield  {author} {\bibinfo {author} {\bibfnamefont {G.}~\bibnamefont
  {Kresse}}\ and\ \bibinfo {author} {\bibfnamefont {D.}~\bibnamefont
  {Joubert}},\ }\bibfield  {title} {\enquote {\bibinfo {title} {From ultrasoft
  pseudopotentials to the projector augmented-wave method},}\ }\href {\doibase
  10.1103/PhysRevB.59.1758} {\bibfield  {journal} {\bibinfo  {journal} {Phys.
  Rev. B}\ }\textbf {\bibinfo {volume} {59}},\ \bibinfo {pages} {1758--1775}
  (\bibinfo {year} {1999})}\BibitemShut {NoStop}%
\bibitem [{\citenamefont {Wang}\ \emph {et~al.}(2017)\citenamefont {Wang},
  \citenamefont {Qi}, \citenamefont {Chen},\ and\ \citenamefont
  {Meng}}]{YCWang2017}%
  \BibitemOpen
  \bibfield  {author} {\bibinfo {author} {\bibfnamefont {Y.-C.}\ \bibnamefont
  {Wang}}, \bibinfo {author} {\bibfnamefont {Y.}~\bibnamefont {Qi}}, \bibinfo
  {author} {\bibfnamefont {S.}~\bibnamefont {Chen}}, \ and\ \bibinfo {author}
  {\bibfnamefont {Z.~Y.}\ \bibnamefont {Meng}},\ }\bibfield  {title} {\enquote
  {\bibinfo {title} {Caution on emergent continuous symmetry: A {Monte Carlo}
  investigation of the transverse-field frustrated {Ising} model on the
  triangular and honeycomb lattices},}\ }\href {\doibase
  10.1103/PhysRevB.96.115160} {\bibfield  {journal} {\bibinfo  {journal} {Phys.
  Rev. B}\ }\textbf {\bibinfo {volume} {96}},\ \bibinfo {pages} {115160}
  (\bibinfo {year} {2017})}\BibitemShut {NoStop}%
\bibitem [{\citenamefont {Powalski}\ \emph {et~al.}(2013)\citenamefont
  {Powalski}, \citenamefont {Coester}, \citenamefont {Moessner},\ and\
  \citenamefont {Schmidt}}]{Powalski2013}%
  \BibitemOpen
  \bibfield  {author} {\bibinfo {author} {\bibfnamefont {M.}~\bibnamefont
  {Powalski}}, \bibinfo {author} {\bibfnamefont {K.}~\bibnamefont {Coester}},
  \bibinfo {author} {\bibfnamefont {R.}~\bibnamefont {Moessner}}, \ and\
  \bibinfo {author} {\bibfnamefont {K.~P.}\ \bibnamefont {Schmidt}},\
  }\bibfield  {title} {\enquote {\bibinfo {title} {Disorder by disorder and
  flat bands in the kagome transverse field ising model},}\ }\href {\doibase
  10.1103/PhysRevB.87.054404} {\bibfield  {journal} {\bibinfo  {journal} {Phys.
  Rev. B}\ }\textbf {\bibinfo {volume} {87}},\ \bibinfo {pages} {054404}
  (\bibinfo {year} {2013})}\BibitemShut {NoStop}%
\bibitem [{\citenamefont {Fey}\ \emph {et~al.}(2019)\citenamefont {Fey},
  \citenamefont {Kapfer},\ and\ \citenamefont {Schmidt}}]{Fey2019}%
  \BibitemOpen
  \bibfield  {author} {\bibinfo {author} {\bibfnamefont {S.}~\bibnamefont
  {Fey}}, \bibinfo {author} {\bibfnamefont {S.~C.}\ \bibnamefont {Kapfer}}, \
  and\ \bibinfo {author} {\bibfnamefont {K.~P.}\ \bibnamefont {Schmidt}},\
  }\bibfield  {title} {\enquote {\bibinfo {title} {Quantum criticality of
  two-dimensional quantum magnets with long-range interactions},}\ }\href
  {\doibase 10.1103/PhysRevLett.122.017203} {\bibfield  {journal} {\bibinfo
  {journal} {Phys. Rev. Lett.}\ }\textbf {\bibinfo {volume} {122}},\ \bibinfo
  {pages} {017203} (\bibinfo {year} {2019})}\BibitemShut {NoStop}%
\bibitem [{\citenamefont {Metcalf}(1974)}]{Metcalf1974}%
  \BibitemOpen
  \bibfield  {author} {\bibinfo {author} {\bibfnamefont {B.~D.}\ \bibnamefont
  {Metcalf}},\ }\bibfield  {title} {\enquote {\bibinfo {title} {Ground state
  spin orderings of the triangular ising model with the nearest and next
  nearest neighbor interaction},}\ }\href@noop {} {\bibfield  {journal}
  {\bibinfo  {journal} {Phys. Lett. A}\ }\textbf {\bibinfo {volume} {46}},\
  \bibinfo {pages} {325--326} (\bibinfo {year} {1974})}\BibitemShut {NoStop}%
\bibitem [{\citenamefont {Wheeler}\ \emph {et~al.}(2009)\citenamefont
  {Wheeler}, \citenamefont {Coldea}, \citenamefont
  {Wawrzy\ifmmode~\acute{n}\else \'{n}\fi{}ska}, \citenamefont {S\"orgel},
  \citenamefont {Jansen}, \citenamefont {Koza}, \citenamefont {Taylor},
  \citenamefont {Adroguer},\ and\ \citenamefont {Shannon}}]{Wheeler2009}%
  \BibitemOpen
  \bibfield  {author} {\bibinfo {author} {\bibfnamefont {E.~M.}\ \bibnamefont
  {Wheeler}}, \bibinfo {author} {\bibfnamefont {R.}~\bibnamefont {Coldea}},
  \bibinfo {author} {\bibfnamefont {E.}~\bibnamefont
  {Wawrzy\ifmmode~\acute{n}\else \'{n}\fi{}ska}}, \bibinfo {author}
  {\bibfnamefont {T.}~\bibnamefont {S\"orgel}}, \bibinfo {author}
  {\bibfnamefont {M.}~\bibnamefont {Jansen}}, \bibinfo {author} {\bibfnamefont
  {M.~M.}\ \bibnamefont {Koza}}, \bibinfo {author} {\bibfnamefont
  {J.}~\bibnamefont {Taylor}}, \bibinfo {author} {\bibfnamefont
  {P.}~\bibnamefont {Adroguer}}, \ and\ \bibinfo {author} {\bibfnamefont
  {N.}~\bibnamefont {Shannon}},\ }\bibfield  {title} {\enquote {\bibinfo
  {title} {Spin dynamics of the frustrated easy-axis triangular antiferromagnet
  {$2H$-AgNiO$_2$} explored by inelastic neutron scattering},}\ }\href
  {\doibase 10.1103/PhysRevB.79.104421} {\bibfield  {journal} {\bibinfo
  {journal} {Phys. Rev. B}\ }\textbf {\bibinfo {volume} {79}},\ \bibinfo
  {pages} {104421} (\bibinfo {year} {2009})}\BibitemShut {NoStop}%
\bibitem [{\citenamefont {Damle}(2015)}]{Damle2015}%
  \BibitemOpen
  \bibfield  {author} {\bibinfo {author} {\bibfnamefont {K.}~\bibnamefont
  {Damle}},\ }\bibfield  {title} {\enquote {\bibinfo {title} {Melting of
  three-sublattice order in easy-axis antiferromagnets on triangular and kagome
  lattices},}\ }\href {\doibase 10.1103/PhysRevLett.115.127204} {\bibfield
  {journal} {\bibinfo  {journal} {Phys. Rev. Lett.}\ }\textbf {\bibinfo
  {volume} {115}},\ \bibinfo {pages} {127204} (\bibinfo {year}
  {2015})}\BibitemShut {NoStop}%
\bibitem [{\citenamefont {Biswas}\ and\ \citenamefont
  {Damle}(2018)}]{Biswas2018}%
  \BibitemOpen
  \bibfield  {author} {\bibinfo {author} {\bibfnamefont {S.}~\bibnamefont
  {Biswas}}\ and\ \bibinfo {author} {\bibfnamefont {K.}~\bibnamefont {Damle}},\
  }\bibfield  {title} {\enquote {\bibinfo {title} {Singular ferromagnetic
  susceptibility of the transverse-field {Ising} antiferromagnet on the
  triangular lattice},}\ }\href {\doibase 10.1103/PhysRevB.97.085114}
  {\bibfield  {journal} {\bibinfo  {journal} {Phys. Rev. B}\ }\textbf {\bibinfo
  {volume} {97}},\ \bibinfo {pages} {085114} (\bibinfo {year}
  {2018})}\BibitemShut {NoStop}%
\bibitem [{\citenamefont {Donnelly}(1974)}]{Donnelly1974}%
  \BibitemOpen
  \bibfield  {author} {\bibinfo {author} {\bibfnamefont {R.~J.}\ \bibnamefont
  {Donnelly}},\ }\enquote {\bibinfo {title} {The ghost of a vanished vortex
  ring},}\ in\ \href {\doibase 10.1007/978-1-4613-4532-9_18} {\emph {\bibinfo
  {booktitle} {Quantum Statistical Mechanics in the Natural Sciences: A Volume
  Dedicated to Lars Onsager on the Occasion of his Seventieth Birthday}}},\
  \bibinfo {editor} {edited by\ \bibinfo {editor} {\bibfnamefont {S.~L.}\
  \bibnamefont {Mintz}}\ and\ \bibinfo {editor} {\bibfnamefont {S.~M.}\
  \bibnamefont {Widmayer}}}\ (\bibinfo  {publisher} {Springer US},\ \bibinfo
  {address} {Boston, MA},\ \bibinfo {year} {1974})\ pp.\ \bibinfo {pages}
  {359--402}\BibitemShut {NoStop}%
\bibitem [{\citenamefont {Nozi{\`e}res}(2004)}]{Nozieres2004}%
  \BibitemOpen
  \bibfield  {author} {\bibinfo {author} {\bibfnamefont {P.}~\bibnamefont
  {Nozi{\`e}res}},\ }\bibfield  {title} {\enquote {\bibinfo {title} {Is the
  roton in superfluid {He$^4$} the ghost of a {Bragg} spot?}}\ }\href {\doibase
  10.1023/B:JOLT.0000044234.82957.2f} {\bibfield  {journal} {\bibinfo
  {journal} {J. Low Temp. Phys.}\ }\textbf {\bibinfo {volume} {137}},\ \bibinfo
  {pages} {45--67} (\bibinfo {year} {2004})}\BibitemShut {NoStop}%
\bibitem [{\citenamefont {Carneiro}\ \emph {et~al.}(1976)\citenamefont
  {Carneiro}, \citenamefont {Ellenson}, \citenamefont {Passell}, \citenamefont
  {McTague},\ and\ \citenamefont {Taub}}]{Carneiro1976}%
  \BibitemOpen
  \bibfield  {author} {\bibinfo {author} {\bibfnamefont {K.}~\bibnamefont
  {Carneiro}}, \bibinfo {author} {\bibfnamefont {W.~D.}\ \bibnamefont
  {Ellenson}}, \bibinfo {author} {\bibfnamefont {L.}~\bibnamefont {Passell}},
  \bibinfo {author} {\bibfnamefont {J.~P.}\ \bibnamefont {McTague}}, \ and\
  \bibinfo {author} {\bibfnamefont {H.}~\bibnamefont {Taub}},\ }\bibfield
  {title} {\enquote {\bibinfo {title} {Neutron-scattering study of the
  structure of adsorbed helium monolayers and of the excitation spectrum of
  few-atomic-layer superfluid films},}\ }\href {\doibase
  10.1103/PhysRevLett.37.1695} {\bibfield  {journal} {\bibinfo  {journal}
  {Phys. Rev. Lett.}\ }\textbf {\bibinfo {volume} {37}},\ \bibinfo {pages}
  {1695--1698} (\bibinfo {year} {1976})}\BibitemShut {NoStop}%
\bibitem [{\citenamefont {Starykh}\ \emph {et~al.}(2006)\citenamefont
  {Starykh}, \citenamefont {Chubukov},\ and\ \citenamefont
  {Abanov}}]{Starykh2006}%
  \BibitemOpen
  \bibfield  {author} {\bibinfo {author} {\bibfnamefont {O.~A.}\ \bibnamefont
  {Starykh}}, \bibinfo {author} {\bibfnamefont {A.~V.}\ \bibnamefont
  {Chubukov}}, \ and\ \bibinfo {author} {\bibfnamefont {A.~G.}\ \bibnamefont
  {Abanov}},\ }\bibfield  {title} {\enquote {\bibinfo {title} {Flat spin-wave
  dispersion in a triangular antiferromagnet},}\ }\href {\doibase
  10.1103/PhysRevB.74.180403} {\bibfield  {journal} {\bibinfo  {journal} {Phys.
  Rev. B}\ }\textbf {\bibinfo {volume} {74}},\ \bibinfo {pages} {180403(R)}
  (\bibinfo {year} {2006})}\BibitemShut {NoStop}%
\bibitem [{\citenamefont {Chernyshev}\ and\ \citenamefont
  {Zhitomirsky}(2006)}]{Chernyshev06}%
  \BibitemOpen
  \bibfield  {author} {\bibinfo {author} {\bibfnamefont {A.~L.}\ \bibnamefont
  {Chernyshev}}\ and\ \bibinfo {author} {\bibfnamefont {M.~E.}\ \bibnamefont
  {Zhitomirsky}},\ }\bibfield  {title} {\enquote {\bibinfo {title} {Magnon
  decay in noncollinear quantum antiferromagnets},}\ }\href {\doibase
  10.1103/PhysRevLett.97.207202} {\bibfield  {journal} {\bibinfo  {journal}
  {Phys. Rev. Lett.}\ }\textbf {\bibinfo {volume} {97}},\ \bibinfo {pages}
  {207202} (\bibinfo {year} {2006})}\BibitemShut {NoStop}%
\bibitem [{\citenamefont {Zheng}\ \emph
  {et~al.}(2006{\natexlab{a}})\citenamefont {Zheng}, \citenamefont
  {Fj\ae{}restad}, \citenamefont {Singh}, \citenamefont {McKenzie},\ and\
  \citenamefont {Coldea}}]{Zheng2006PRL}%
  \BibitemOpen
  \bibfield  {author} {\bibinfo {author} {\bibfnamefont {Weihong}\ \bibnamefont
  {Zheng}}, \bibinfo {author} {\bibfnamefont {John~O.}\ \bibnamefont
  {Fj\ae{}restad}}, \bibinfo {author} {\bibfnamefont {Rajiv R.~P.}\
  \bibnamefont {Singh}}, \bibinfo {author} {\bibfnamefont {Ross~H.}\
  \bibnamefont {McKenzie}}, \ and\ \bibinfo {author} {\bibfnamefont {Radu}\
  \bibnamefont {Coldea}},\ }\bibfield  {title} {\enquote {\bibinfo {title}
  {Anomalous excitation spectra of frustrated quantum antiferromagnets},}\
  }\href {\doibase 10.1103/PhysRevLett.96.057201} {\bibfield  {journal}
  {\bibinfo  {journal} {Phys. Rev. Lett.}\ }\textbf {\bibinfo {volume} {96}},\
  \bibinfo {pages} {057201} (\bibinfo {year} {2006}{\natexlab{a}})}\BibitemShut
  {NoStop}%
\bibitem [{\citenamefont {Zheng}\ \emph
  {et~al.}(2006{\natexlab{b}})\citenamefont {Zheng}, \citenamefont
  {Fj\ae{}restad}, \citenamefont {Singh}, \citenamefont {McKenzie},\ and\
  \citenamefont {Coldea}}]{Zheng2006PRB}%
  \BibitemOpen
  \bibfield  {author} {\bibinfo {author} {\bibfnamefont {Weihong}\ \bibnamefont
  {Zheng}}, \bibinfo {author} {\bibfnamefont {John~O.}\ \bibnamefont
  {Fj\ae{}restad}}, \bibinfo {author} {\bibfnamefont {Rajiv R.~P.}\
  \bibnamefont {Singh}}, \bibinfo {author} {\bibfnamefont {Ross~H.}\
  \bibnamefont {McKenzie}}, \ and\ \bibinfo {author} {\bibfnamefont {Radu}\
  \bibnamefont {Coldea}},\ }\bibfield  {title} {\enquote {\bibinfo {title}
  {Excitation spectra of the spin-$\frac{1}{2}$ triangular-lattice {Heisenberg}
  antiferromagnet},}\ }\href {\doibase 10.1103/PhysRevB.74.224420} {\bibfield
  {journal} {\bibinfo  {journal} {Phys. Rev. B}\ }\textbf {\bibinfo {volume}
  {74}},\ \bibinfo {pages} {224420} (\bibinfo {year}
  {2006}{\natexlab{b}})}\BibitemShut {NoStop}%
\bibitem [{\citenamefont {Verresen}\ \emph {et~al.}(2019)\citenamefont
  {Verresen}, \citenamefont {Moessner},\ and\ \citenamefont
  {Pollmann}}]{Verresen2019}%
  \BibitemOpen
  \bibfield  {author} {\bibinfo {author} {\bibfnamefont {Ruben}\ \bibnamefont
  {Verresen}}, \bibinfo {author} {\bibfnamefont {Roderich}\ \bibnamefont
  {Moessner}}, \ and\ \bibinfo {author} {\bibfnamefont {Frank}\ \bibnamefont
  {Pollmann}},\ }\bibfield  {title} {\enquote {\bibinfo {title} {Avoided
  quasiparticle decay from strong quantum interactions},}\ }\href {\doibase
  10.1038/s41567-019-0535-3} {\bibfield  {journal} {\bibinfo  {journal} {Nat.
  Phys.}\ }\textbf {\bibinfo {volume} {15}},\ \bibinfo {pages} {750--753}
  (\bibinfo {year} {2019})}\BibitemShut {NoStop}%
\bibitem [{\citenamefont {Ferrari}\ and\ \citenamefont
  {Becca}(2019)}]{Farrari2019}%
  \BibitemOpen
  \bibfield  {author} {\bibinfo {author} {\bibfnamefont {Francesco}\
  \bibnamefont {Ferrari}}\ and\ \bibinfo {author} {\bibfnamefont {Federico}\
  \bibnamefont {Becca}},\ }\bibfield  {title} {\enquote {\bibinfo {title}
  {Dynamical structure factor of the ${J}_{1}\ensuremath{-}{J}_{2}$ heisenberg
  model on the triangular lattice: Magnons, spinons, and gauge fields},}\
  }\href {\doibase 10.1103/PhysRevX.9.031026} {\bibfield  {journal} {\bibinfo
  {journal} {Phys. Rev. X}\ }\textbf {\bibinfo {volume} {9}},\ \bibinfo {pages}
  {031026} (\bibinfo {year} {2019})}\BibitemShut {NoStop}%
\bibitem [{\citenamefont {Chen}\ \emph {et~al.}(2019)\citenamefont {Chen},
  \citenamefont {Qu}, \citenamefont {Li}, \citenamefont {Chen}, \citenamefont
  {Gong}, \citenamefont {von Delft}, \citenamefont {Weichselbaum},\ and\
  \citenamefont {Li}}]{Chen2019}%
  \BibitemOpen
  \bibfield  {author} {\bibinfo {author} {\bibfnamefont {L.}~\bibnamefont
  {Chen}}, \bibinfo {author} {\bibfnamefont {D.-W.}\ \bibnamefont {Qu}},
  \bibinfo {author} {\bibfnamefont {H.}~\bibnamefont {Li}}, \bibinfo {author}
  {\bibfnamefont {B.-B.}\ \bibnamefont {Chen}}, \bibinfo {author}
  {\bibfnamefont {S.-S.}\ \bibnamefont {Gong}}, \bibinfo {author}
  {\bibfnamefont {J.}~\bibnamefont {von Delft}}, \bibinfo {author}
  {\bibfnamefont {A.}~\bibnamefont {Weichselbaum}}, \ and\ \bibinfo {author}
  {\bibfnamefont {W.}~\bibnamefont {Li}},\ }\bibfield  {title} {\enquote
  {\bibinfo {title} {Two temperature scales in the triangular lattice
  {Heisenberg} antiferromagnet},}\ }\href {\doibase 10.1103/PhysRevB.99.140404}
  {\bibfield  {journal} {\bibinfo  {journal} {Phys. Rev. B}\ }\textbf {\bibinfo
  {volume} {99}},\ \bibinfo {pages} {140404(R)} (\bibinfo {year}
  {2019})}\BibitemShut {NoStop}%
\bibitem [{\citenamefont {Minnhagen}(1987)}]{Minnhagen1987}%
  \BibitemOpen
  \bibfield  {author} {\bibinfo {author} {\bibfnamefont {P.}~\bibnamefont
  {Minnhagen}},\ }\bibfield  {title} {\enquote {\bibinfo {title} {The
  two-dimensional {Coulomb} gas, vortex unbinding, and
  superfluid-superconducting films},}\ }\href {\doibase
  10.1103/RevModPhys.59.1001} {\bibfield  {journal} {\bibinfo  {journal} {Rev.
  Mod. Phys.}\ }\textbf {\bibinfo {volume} {59}},\ \bibinfo {pages}
  {1001--1066} (\bibinfo {year} {1987})}\BibitemShut {NoStop}%
\bibitem [{\citenamefont {Banerjee}\ \emph {et~al.}(2016)\citenamefont
  {Banerjee}, \citenamefont {Bridges}, \citenamefont {Yan}, \citenamefont
  {Aczel}, \citenamefont {Li}, \citenamefont {Stone}, \citenamefont {Granroth},
  \citenamefont {Lumsden}, \citenamefont {Yiu}, \citenamefont {Knolle},
  \citenamefont {Bhattacharjee}, \citenamefont {Kovrizhin}, \citenamefont
  {Moessner}, \citenamefont {Tennant}, \citenamefont {Mandrus},\ and\
  \citenamefont {Nagler}}]{Banerjee2016}%
  \BibitemOpen
  \bibfield  {author} {\bibinfo {author} {\bibfnamefont {A.}~\bibnamefont
  {Banerjee}}, \bibinfo {author} {\bibfnamefont {C.~A.}\ \bibnamefont
  {Bridges}}, \bibinfo {author} {\bibfnamefont {J.~Q.}\ \bibnamefont {Yan}},
  \bibinfo {author} {\bibfnamefont {A.~A.}\ \bibnamefont {Aczel}}, \bibinfo
  {author} {\bibfnamefont {L.}~\bibnamefont {Li}}, \bibinfo {author}
  {\bibfnamefont {M.~B.}\ \bibnamefont {Stone}}, \bibinfo {author}
  {\bibfnamefont {G.~E.}\ \bibnamefont {Granroth}}, \bibinfo {author}
  {\bibfnamefont {M.~D.}\ \bibnamefont {Lumsden}}, \bibinfo {author}
  {\bibfnamefont {Y.}~\bibnamefont {Yiu}}, \bibinfo {author} {\bibfnamefont
  {J.}~\bibnamefont {Knolle}}, \bibinfo {author} {\bibfnamefont
  {S.}~\bibnamefont {Bhattacharjee}}, \bibinfo {author} {\bibfnamefont {D.~L.}\
  \bibnamefont {Kovrizhin}}, \bibinfo {author} {\bibfnamefont {R.}~\bibnamefont
  {Moessner}}, \bibinfo {author} {\bibfnamefont {D.~A.}\ \bibnamefont
  {Tennant}}, \bibinfo {author} {\bibfnamefont {D.~G.}\ \bibnamefont
  {Mandrus}}, \ and\ \bibinfo {author} {\bibfnamefont {S.~E.}\ \bibnamefont
  {Nagler}},\ }\bibfield  {title} {\enquote {\bibinfo {title} {Proximate kitaev
  quantum spin liquid behaviour in a honeycomb magnet},}\ }\href {\doibase
  10.1038/nmat4604} {\bibfield  {journal} {\bibinfo  {journal} {Nat. Mater.}\
  }\textbf {\bibinfo {volume} {15}},\ \bibinfo {pages} {733--740} (\bibinfo
  {year} {2016})}\BibitemShut {NoStop}%
\bibitem [{\citenamefont {Lake}\ \emph {et~al.}(2005)\citenamefont {Lake},
  \citenamefont {Tennant}, \citenamefont {Frost},\ and\ \citenamefont
  {Nagler}}]{Lake2005}%
  \BibitemOpen
  \bibfield  {author} {\bibinfo {author} {\bibfnamefont {B.}~\bibnamefont
  {Lake}}, \bibinfo {author} {\bibfnamefont {D.~A.}\ \bibnamefont {Tennant}},
  \bibinfo {author} {\bibfnamefont {C.~D.}\ \bibnamefont {Frost}}, \ and\
  \bibinfo {author} {\bibfnamefont {S.~E.}\ \bibnamefont {Nagler}},\ }\bibfield
   {title} {\enquote {\bibinfo {title} {Quantum criticality and universal
  scaling of a quantum antiferromagnet},}\ }\href {\doibase 10.1038/nmat1327}
  {\bibfield  {journal} {\bibinfo  {journal} {Nat. Mater.}\ }\textbf {\bibinfo
  {volume} {4}},\ \bibinfo {pages} {329--334} (\bibinfo {year}
  {2005})}\BibitemShut {NoStop}%
\bibitem [{\citenamefont {Nelson}\ and\ \citenamefont
  {Kosterlitz}(1977)}]{Nelson1977}%
  \BibitemOpen
  \bibfield  {author} {\bibinfo {author} {\bibfnamefont {D.~R.}\ \bibnamefont
  {Nelson}}\ and\ \bibinfo {author} {\bibfnamefont {J.~M.}\ \bibnamefont
  {Kosterlitz}},\ }\bibfield  {title} {\enquote {\bibinfo {title} {Universal
  jump in the superfluid density of two-dimensional superfluids},}\ }\href
  {\doibase 10.1103/PhysRevLett.39.1201} {\bibfield  {journal} {\bibinfo
  {journal} {Phys. Rev. Lett.}\ }\textbf {\bibinfo {volume} {39}},\ \bibinfo
  {pages} {1201--1205} (\bibinfo {year} {1977})}\BibitemShut {NoStop}%
\bibitem [{\citenamefont {Chatelain}()}]{Chatelain2014}%
  \BibitemOpen
  \bibfield  {author} {\bibinfo {author} {\bibfnamefont {C.}~\bibnamefont
  {Chatelain}},\ }\bibfield  {title} {\enquote {\bibinfo {title} {{DMRG} study
  of the {Berezinskii{\textendash}Kosterlitz{\textendash}Thouless} transitions
  of the {2D} five-state clock model},}\ }\href {\doibase
  10.1088/1742-5468/2014/11/p11022} {\bibfield  {journal} {\bibinfo  {journal}
  {J. Stat. Mech.}\ }\textbf {\bibinfo {volume} {(2014)}},\ \bibinfo {pages}
  {P11022}}\BibitemShut {NoStop}%
\bibitem [{\citenamefont {Chen}\ \emph {et~al.}(2017)\citenamefont {Chen},
  \citenamefont {Liu}, \citenamefont {Chen},\ and\ \citenamefont
  {Li}}]{Chen.b+:2017:SETTN}%
  \BibitemOpen
  \bibfield  {author} {\bibinfo {author} {\bibfnamefont {B.-B.}\ \bibnamefont
  {Chen}}, \bibinfo {author} {\bibfnamefont {Y.-J.}\ \bibnamefont {Liu}},
  \bibinfo {author} {\bibfnamefont {Z.}~\bibnamefont {Chen}}, \ and\ \bibinfo
  {author} {\bibfnamefont {W.}~\bibnamefont {Li}},\ }\bibfield  {title}
  {\enquote {\bibinfo {title} {Series-expansion thermal tensor network approach
  for quantum lattice models},}\ }\href {\doibase 10.1103/PhysRevB.95.161104}
  {\bibfield  {journal} {\bibinfo  {journal} {Phys. Rev. B}\ }\textbf {\bibinfo
  {volume} {95}},\ \bibinfo {pages} {161104(R)} (\bibinfo {year}
  {2017})}\BibitemShut {NoStop}%
\bibitem [{\citenamefont {Li}\ \emph {et~al.}(2011)\citenamefont {Li},
  \citenamefont {Ran}, \citenamefont {Gong}, \citenamefont {Zhao},
  \citenamefont {Xi}, \citenamefont {Ye},\ and\ \citenamefont
  {Su}}]{Li.w+:2011:LTRG}%
  \BibitemOpen
  \bibfield  {author} {\bibinfo {author} {\bibfnamefont {W.}~\bibnamefont
  {Li}}, \bibinfo {author} {\bibfnamefont {S.-J.}\ \bibnamefont {Ran}},
  \bibinfo {author} {\bibfnamefont {S.-S.}\ \bibnamefont {Gong}}, \bibinfo
  {author} {\bibfnamefont {Y.}~\bibnamefont {Zhao}}, \bibinfo {author}
  {\bibfnamefont {B.}~\bibnamefont {Xi}}, \bibinfo {author} {\bibfnamefont
  {F.}~\bibnamefont {Ye}}, \ and\ \bibinfo {author} {\bibfnamefont
  {G.}~\bibnamefont {Su}},\ }\bibfield  {title} {\enquote {\bibinfo {title}
  {Linearized tensor renormalization group algorithm for the calculation of
  thermodynamic properties of quantum lattice models},}\ }\href {\doibase
  10.1103/PhysRevLett.106.127202} {\bibfield  {journal} {\bibinfo  {journal}
  {Phys. Rev. Lett.}\ }\textbf {\bibinfo {volume} {106}},\ \bibinfo {pages}
  {127202} (\bibinfo {year} {2011})}\BibitemShut {NoStop}%
\bibitem [{\citenamefont {Li}\ \emph {et~al.}(2019{\natexlab{b}})\citenamefont
  {Li}, \citenamefont {Chen}, \citenamefont {Chen}, \citenamefont {von Delft},
  \citenamefont {Weichselbaum},\ and\ \citenamefont {Li}}]{Lih2019}%
  \BibitemOpen
  \bibfield  {author} {\bibinfo {author} {\bibfnamefont {H.}~\bibnamefont
  {Li}}, \bibinfo {author} {\bibfnamefont {B.-B.}\ \bibnamefont {Chen}},
  \bibinfo {author} {\bibfnamefont {Z.}~\bibnamefont {Chen}}, \bibinfo {author}
  {\bibfnamefont {J.}~\bibnamefont {von Delft}}, \bibinfo {author}
  {\bibfnamefont {A.}~\bibnamefont {Weichselbaum}}, \ and\ \bibinfo {author}
  {\bibfnamefont {W.}~\bibnamefont {Li}},\ }\bibfield  {title} {\enquote
  {\bibinfo {title} {Thermal tensor renormalization group simulations of
  square-lattice quantum spin models},}\ }\href {\doibase
  10.1103/PhysRevB.100.045110} {\bibfield  {journal} {\bibinfo  {journal}
  {Phys. Rev. B}\ }\textbf {\bibinfo {volume} {100}},\ \bibinfo {pages}
  {045110} (\bibinfo {year} {2019}{\natexlab{b}})}\BibitemShut {NoStop}%
\end{thebibliography}%

\section*{Acknowledgements}
The authors are indebted to Ziyu Chen, Wentao Jin, Andreas Weichselbaum, Yao Shen, and Yue-Sheng Li for stimulating discussions.
This work was supported by the Ministry of Science and Technology of China through the National Key Research and Development Program (Grant No. 2016YFA0300502), the Strategic Priority Research Program of the Chinese Academy of Sciences (Grant No. XDB28000000), the National Science Foundation of China (Grant No. 11421092, 11574359, 11674370, 11974036, 11874115, and 11834014) and Research Grants Council of Hong Kong Special Administrative Region of China through 17303019 and the Aspen Center for Physics, which is supported by National Science Foundation grant PHY-1607611. B.-B.C. was supported by the German Research foundation, DFG WE4819/3-1. W.L. was supported by the Fundamental Research Funds for the Central Universities. We thank the Center for Quantum Simulation Sciences at Institute of Physics, Chinese Academy of Sciences, the Computational Initiative at the
Faculty of Science at the University of Hong Kong and
the Tianhe-1A platform at the National Supercomputer
Center in Tianjin and Tianhe-2 platform at the National
Supercomputer Center in Guangzhou for their technical
support and generous allocation of CPU time.

\section*{Author contributions}
H.L. and Y.D.L. contributed equally to this work. W.L., Y.Q. and Z.Y.M initiated the work. H.L. and B.B.C. performed the calculations of thermodynamics, X.T.Z. and X.L.S. performed first-principle calculations of the band structure, and Y.D.L. performed the dynamical property simulations. All authors contributed to the analysis of the results. Z.Y.M. and W.L. supervised the project.

\section*{Additional information}
\noindent
\textbf{Supplementary Information} is available in the online version of the paper. \\
\noindent
\textbf{Competing interests:} The authors declare no competing interests. \\

\newpage
\setcounter{equation}{0}
\setcounter{figure}{0}
\setcounter{table}{0}
\setcounter{subsection}{0}

\renewcommand{\theequation}{S\arabic{equation}}
\renewcommand{\thefigure}{S\arabic{figure}}
\renewcommand{\bibnumfmt}[1]{[#1]}
\renewcommand{\citenumfont}[1]{#1}
\renewcommand{\section}

\begin{center}
\section{Supplementary Information for: \\
\textbf{\large{Kosterlitz-Thouless Melting of Magnetic Order in the Triangular Quantum Ising Material TmMgGaO$_4$}}}
\end{center}

\textbf{Supplementary Note 1: The electron density distributions and magnetic interactions in TMGO.}

Density-functional theory (DFT) calculations of TMGO can be performed via the Vienna $ab$ $initio$ simulation package, 
with the projector augmented wave method \cite{VASP1,VASP2}. 
In the DFT calculations, 
we take the TMGO lattice parameters $a = b = 3.4260$ \AA \, and $c =  25.1690$ \AA \, as determined from experiments.

When Coulomb interaction is switched off ($U=0$), the band structure of TMGO indicates a metallic state [Supplementary Fig.~\ref{figS-CHG}(a)], 
while a finite $U$ ($=3$ eV) opens up a Mott gap as shown in Supplementary Fig.~\ref{figS-CHG}(b). 
The partial electron density relevant for magnetic exchange interactions include the contributions 
from 4f electrons of Tm$^{3+}$ and 2p electrons of O$^{2-}$, 
as seen in the density of states in Supplementary Figs.~\ref{figS-CHG}(a,b). 
The Tm-O-Tm superexchange paths are visualized as electron clouds overlap in Supplementary Fig.~\ref{figS-CHG}(c), 
and the Tm$^{3+}$ ions form a triangular lattice [lower plot in Supplementary Fig.~\ref{figS-CHG}(c)]. 
The two dimensionality of the material TMGO is manifested, as the electron density at the interlayer regime 
is negligibly small compared to that in the Tm-O-Tm path. \\

\begin{figure}[h!]
\includegraphics[angle=0,width=1\linewidth]{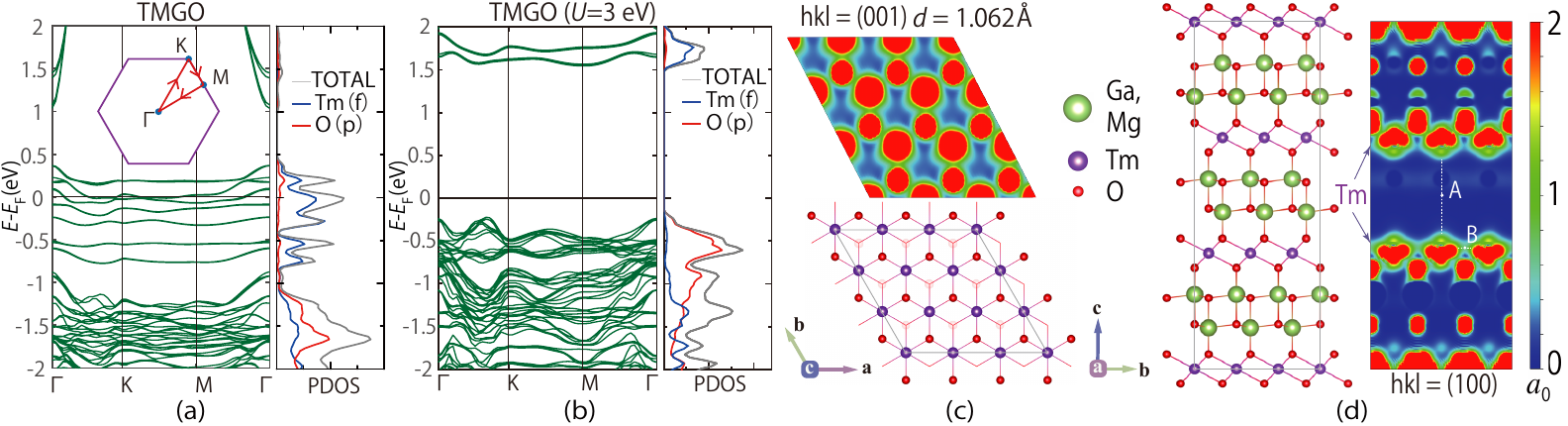}
\renewcommand{\figurename}{\textbf{Supplementary Fig.}}
\caption{\textbf{The crystal structure and partial electron density distributions.}  The energy band results are plotted with (a) $U=0$ and (b) $U=3$ eV, where the corresponding densities of states (total, 4f, and 2p) are also shown. (c) depicts the layered lattice structure and corresponding electron density of TMGO [in the unit $a_{\rm{0}}=10^{-6}$ e$\cdot$Bohr$^{-3}$, and note the density plot (above) corresponds precisely to the regime enclosed by the black diamond-shape box in the lattice structure (below)]. The partial electron density $\rho_{\rm{e}}$ with energies between 1.5 and 2 eV (in the $U=3$ eV calculations) includes mainly 4f and 2p electron contributions. (d) provides a side view of partial electron density $\rho_{\rm{e}}$, where we find $\rho_{\rm{e}}$ at point $A$ (between two layers) and $B$ [a typical point in the superexchange path within the (001) plane] are different by 2-3 orders of magnitude, manifesting the two dimensionality of magnetic couplings in TMGO.}
\label{figS-CHG}
\end{figure}

\textbf{Supplementary Note 2: Spin ordering and phase diagram of the $J_1$-$J_2$ TLI.}

\begin{figure}[t!]
\includegraphics[angle=0,width=0.5\linewidth]{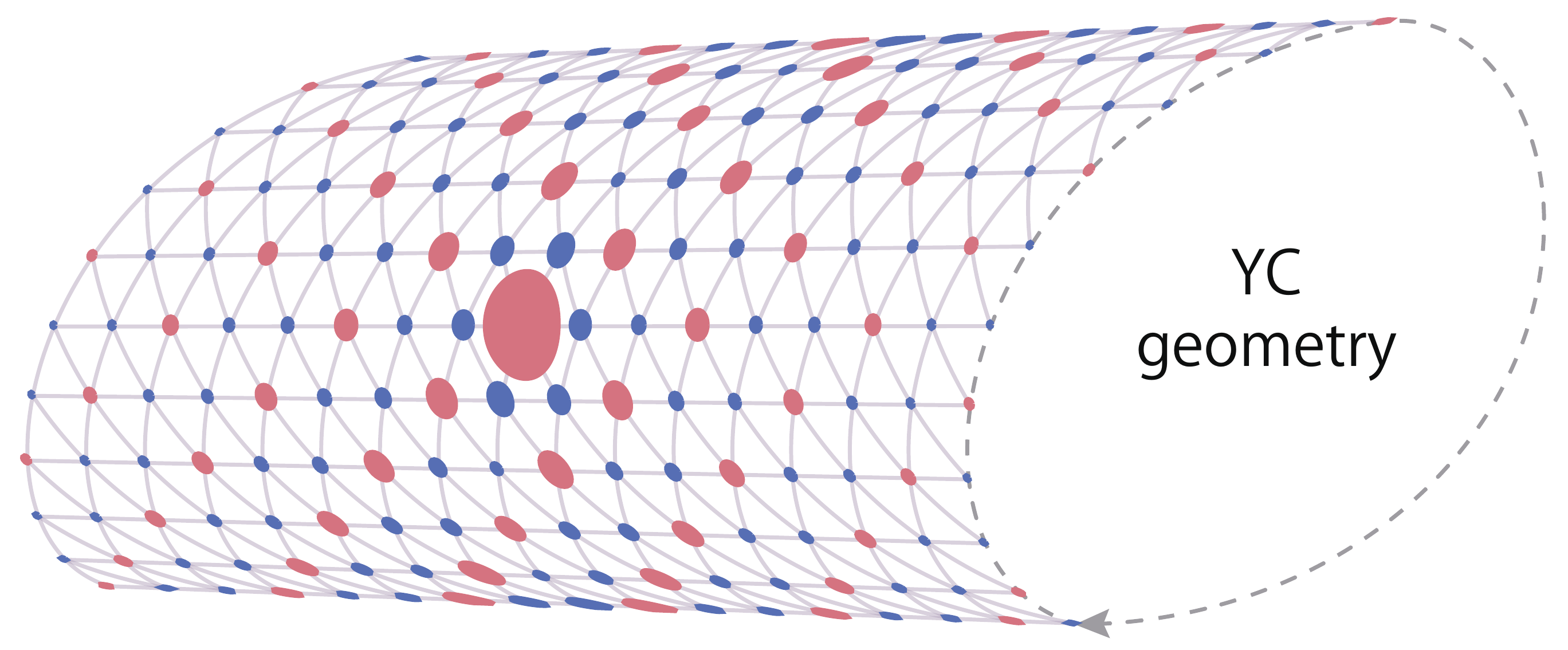}
\renewcommand{\figurename}{\textbf{Supplementary Fig.}}
\caption{\textbf{Real-space correlation $\langle S^z_{\rm{0}} \cdot S^z_{\rm{r}} \rangle$ and a three-sublattice order.} Each solid circle represents the 
real-space $\langle S^z_0 \cdot S^z_{\rm{r}} \rangle$ correlation related to the central site 0, the size of circle denotes the magnitude of correlation, 
and the red(blue) color for the positive(negative) sign. The clock order pattern with an enlarged unit cell can be clearly seen. 
The XTRG calculation is performed with parameters $J_1=0.99$ meV, $J_2=0.05J_1$, and $\Delta=0.54J_1$ at $T\simeq 0.57$ K, 
on the YC geometry also specified in the plot.}
\label{figSzSz}
\end{figure}

\begin{figure}[t!]
\includegraphics[angle=0,width=0.7\linewidth]{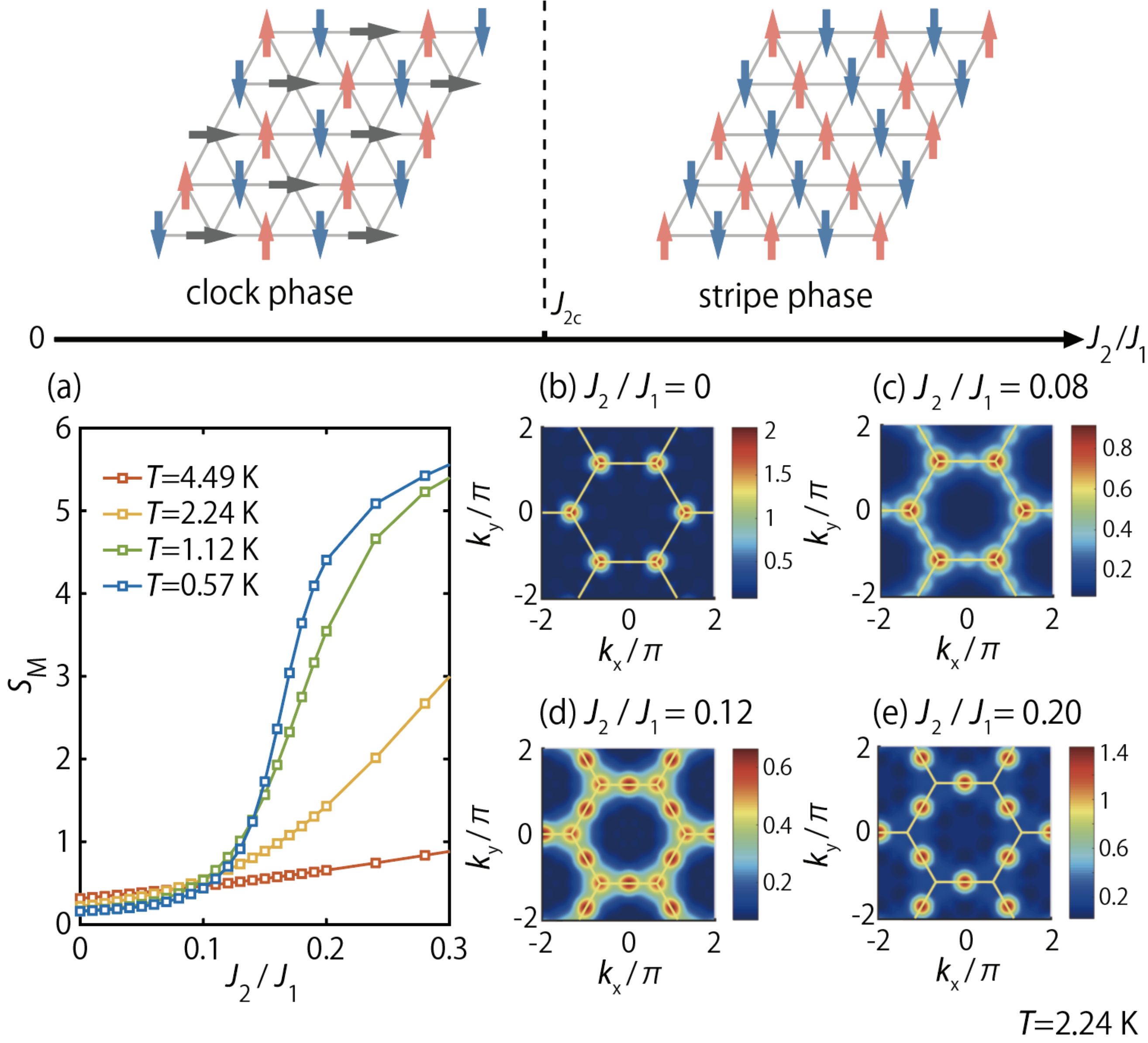}
\renewcommand{\figurename}{\textbf{Supplementary Fig.}}
\caption{\textbf{Schematic phase diagram of the $J_1$-$J_2$ TLI and static structure factors.} As $J_2$ increases, 
there exists a quantum phase transition between the clock and stripe phases, 
taking place at $J_{\rm{2c}}/J_1 \sim 0.1$ (for $\Delta/J_1=0.54$). 
In (a) we collect the M-point intensity and plot it vs. $J_2/J_1$ at four different temperatures, 
from which we see clearly that $S_{\rm{M}}$ in small $J_2$ regime is continuously connected to that 
in the relatively large $J_2$ regime, i.e., the stripy phase. 
The contour plots of $S(q)$ are shown in panels (b-e), 
with various $J_2/J_1$ values (0 to 0.2). In the calculations, 
we fix the parameter as $\Delta/J_1=0.54$, $J_1=0.99$ meV, 
and compute $S(\bf{q})$ at $T$ = 2.24 K for (b-e).}
\label{figS2}
\end{figure}

\begin{figure}[t!]
   \includegraphics[angle=0,width=0.75\linewidth]{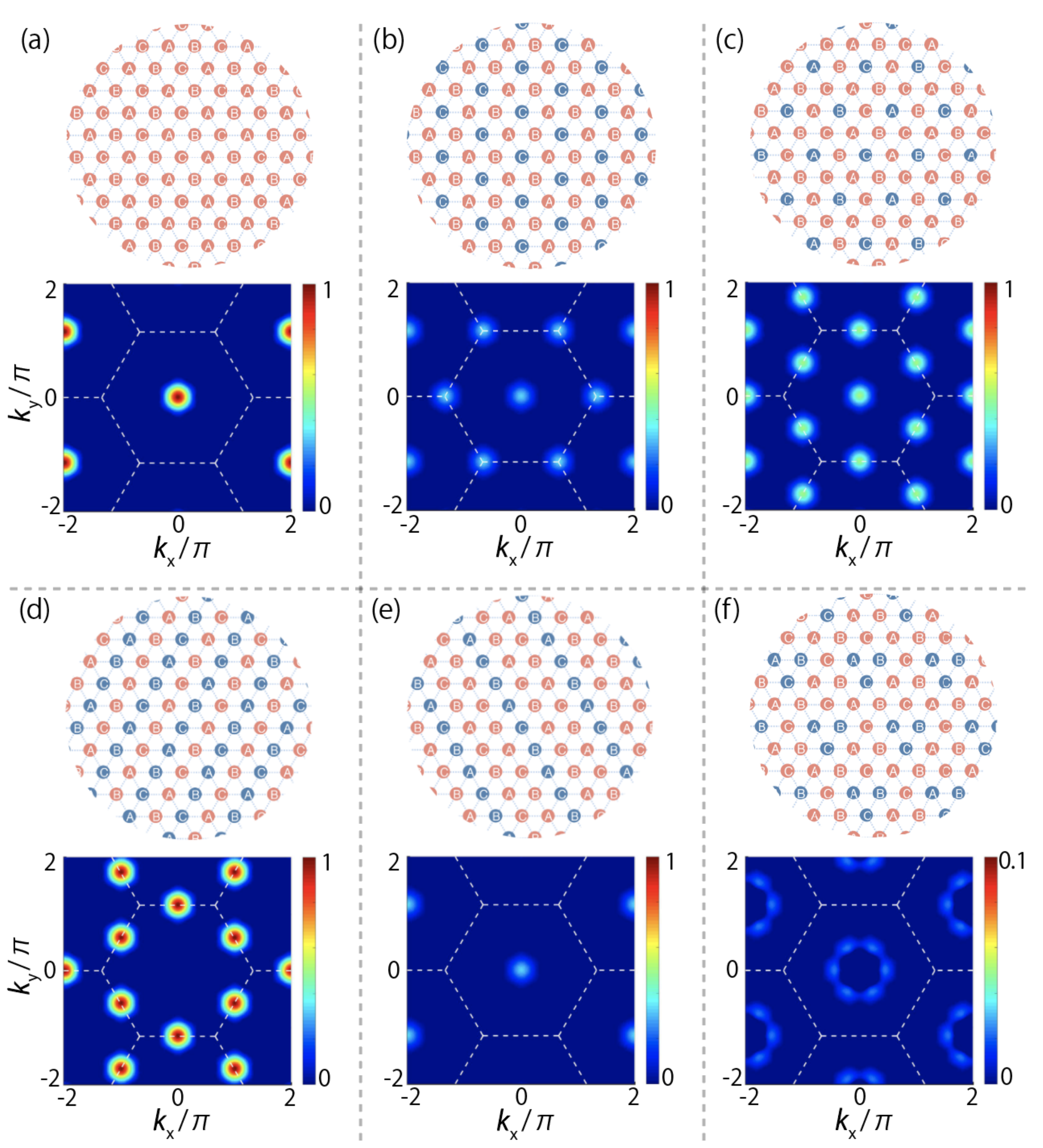}
   \renewcommand{\figurename}{\textbf{Supplementary Fig.}}
    \caption{\textbf{Classical orders of the $J_1-J_2$ TLI model and their static structure factor $S(\mathbf{q})$.} 
    Panels (a-f) plot six different magnetic orders (all possible classical spin configurations from Supplementary Ref.~\cite{Metcalf1974}), 
    along with the corresponding spin structures in reciprocal space. 
    Note the $S(\mathbf{q})$ data are computed in a $12\times12$ cluster, symmetrized, and normalized by the number of sites.}
\label{Fig:cTLI}
\end{figure}

As mentioned in the main text, by increasing the ratio of $J_2/J_1$ we can drive the system from the clock order phase to a stripe order phase. 
To verify this, firstly we calculate the static magnetic structure factor $S(\mathbf{q})$ from $\langle S_0^z S_{\rm{r}}^z \rangle$ correlations,
and show the results in Supplementary Fig.~\ref{figSzSz}. From the real-space correlations on YC6 geometry, 
a three-sublattice order can be clearly identified, which translates into the $S(\mathbf{q})$ peak at the K point 
[see, e.g., Fig.~\B{5}(b) in the main text, or Supplementary Fig.~\ref{figS2}], signifying the presence of clock order. 

However, as $J_2/J_1$ increases to 0.2, $S(\mathbf{q})$ changes its peak to the M point 
[Supplementary Fig.~\ref{figS2}(e)], consistent with a two-sublattice stripe-order pattern. 
Therefore, there must be a quantum phase transition between the clock and stripe phases, probably of first order. 
As the stripe order (with structure factor peak at M point) is in close proximity to the clock phase (a small $J_2$ drives the phase transition), 
we relate the finite energy M-roton excitations in the dynamic spin spectra with instability towards the stripe order.\\

\textbf{Supplementary Note 3: Classical spin orders and their static structure factors.}

The classical spin orders of $J_1-J_2$ TLI has been throughly investigated in Supplementary Ref.~\cite{Metcalf1974}. Here we replot the possible classical spin configurations in Supplementary Fig.~\ref{Fig:cTLI}, along with their computed static structure factor $S(\mathbf{q})$. One can see clearly that only orders in Supplementary Figs.~\ref{Fig:cTLI}(c,d) have peaks at M point. Supplementary Fig.~\ref{Fig:cTLI}(d) corresponds to the stripy order, in agreement with our simulated data (with large $J_2$), while the one in (c) has a $\Gamma$ peak that is absent in our results. Therefore, the only classical Ising configuration that corresponds to our structure factor data is Supplementary Fig.~\ref{Fig:cTLI}(d), i.e., stripy order.

The other way around, in the context of our TLI model, Bragg peak at M $= (1/2,1/2)$ corresponds to a $\pi$ phase shift, i.e., antiferromagnetic correlation along primitive vectors $\bold{a}$ and $\bold{b}$ (see primitive vector in Fig.~\B{1} of the main text), while the correlation is ferromagnetically along $\bold{a}+\bold{b}$ and $\bold{a}-\bold{b}$. This clearly corresponds to a magnetic stripy phase, in the context of TLI model.

Furthermore, we can also argue that large $J_2>0$  favors a stripe order: In the presence of $J_2$, the stripe Ising configuration [Supplementary Fig.~\ref{Fig:cTLI}(d)] leads to an energy estimate of $-J_1-J_2$, while the up-up-down (UUD) order [A,B sublattice spin up and C sublattice down, see Supplementary Fig.~\ref{Fig:cTLI}(b)] $-J_1+3J_2$, and the order in Supplementary Fig.~\ref{Fig:cTLI}(c) $0$ energy. Therefore, it is clear that the stripy configuration is energetically more favorable in large $J_2$ limit. Note that in a recent study of triangular lattice  magnet AgNiO$_2$, the existence of stripe order was observed, which is ascribed to the relatively large $J_2$ in the compound \cite{Wheeler2009}.

\textbf{Supplementary Note 4: TLI parameter fittings.}

The parameter fitting workflow is as follows: we scan the parameters $(J_1, J_2, \Delta)$ to fit the specific heat $C_{\rm{m}}(T)$, magnetic entropy $S_{\rm{m}}(T)$, as well as the susceptibility $\chi(T)$ (at a small magnetic field $h$=1 kOe), and find the optimal parameters. Given that, we compute the magnetization curves (at different $T$) and magnetic entropy $S_{\rm{m}}$ at finite fields $h$, and compare directly to experimental data \cite{Li2018, Shen2018}, so as to ensure that the parameter set is adequate and precise to model the material.

\begin{figure}[t!]
   \includegraphics[angle=0,width=0.75\linewidth]{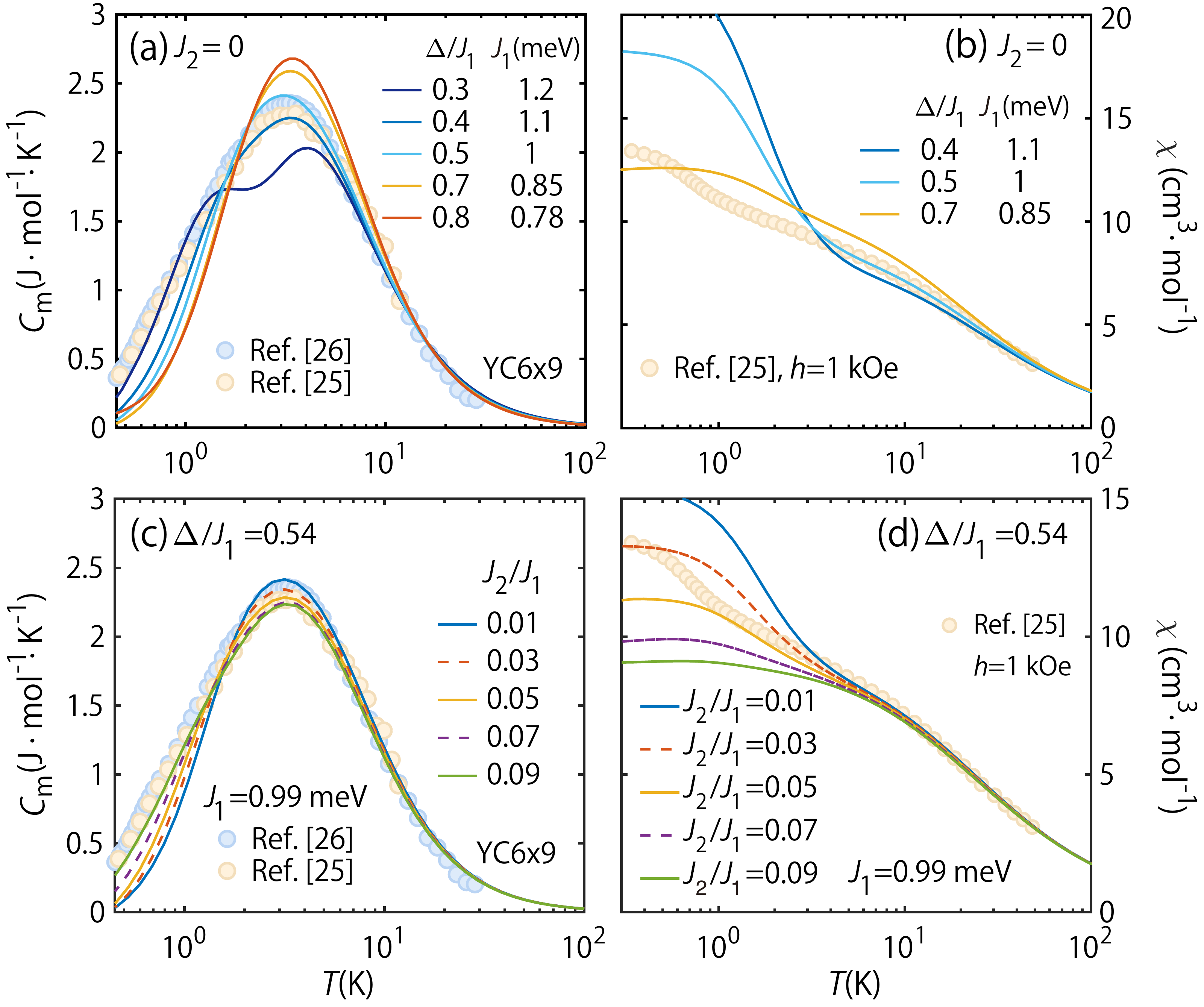}
   \renewcommand{\figurename}{\textbf{Supplementary Fig.}}
    \caption{\textbf{Parameter fittings.} (a) The specific heat $C_{\rm{m}}(T)$ curves with various $\Delta/J_1$ and fixed $J_2$ = 0, obtained by XTRG simulations on the YC6$\times$9 lattice. In panel (b) we show the susceptibility $\chi (T)$ with various $\Delta/J_1$ = 0.4, 0.5, 0.7 (at $h$=1 kOe), which all clearly failed to fit the experimental data well. (c) shows the $C_{\rm{m}} (T)$ results with $J_2/J_1$ ranging from 0.01 to 0.09 and a fixed (optimal) $\Delta/J_1$=0.54, and (d) depicts the susceptibility data correspondingly. $g_\parallel =13.212$ is fixed throughout the fittings (see Supplementary Note 3).}
\label{Fig:ScanParam}
\end{figure}

To be concrete, we show in Supplementary Fig.~\ref{Fig:ScanParam} part of simulation data in our scanning. In Supplementary Fig.~\ref{Fig:ScanParam}(a), we start with $J_2=0$ and scan various $\Delta$ values. It is found that $C_{\rm{m}}$ curves are sensitive, in terms of the peak height as well as the overall shape, to different $\Delta$ values. By tuning $\Delta$ (while keeping $J_2=0$), we find $\Delta/J_1=0.4$ and $J_1=1.1$ meV can produce results in agreement with experimental $C_{\rm{m}}$ curves. However, with this set of parameters (as well as other $\Delta$ values) we clearly miss the experimental susceptibility line, as plotted in Supplementary Fig.~\ref{Fig:ScanParam}(b). It therefore suggests that a finite $J_2$ should be involved in the fittings.

After a thorough scanning in the parameter space, we find $\Delta/J_1=0.54$ with  $J_2/J_1=0.05$ can well reproduce both the specific heat curve in Supplementary Fig.~\ref{Fig:ScanParam}(c) and the magnetic susceptibility data in (d). To show how sensitive the fittings are with respect to $J_2$, we also provide in Supplementary Figs.~\ref{Fig:ScanParam}(c,d) the simulated data with $J_2 = 0.01$ to 0.09, from which we see that $J_2$ considerably influences both $C_m(T)$ and $\chi(T)$ curves. Regarding error bar of the fitted parameter $J_2$, the most probable regime where $J_2/J_1$ resides is between 0.03 and 0.05 (while 0.05 is still more preferable).

With this parameter set  $J_1 = 0.99$ meV, $\Delta/J_1=0.54$, and $J_2/J_1=0.05$ (as well as $g_\parallel =13.212$), we have computed the magnetization curves at two different temperatures and entropy $S_{\rm{m}}$ at finite magnetic fields. We compare them to experimental data in Fig.~\B{2} of the main text and observe excellent agreement. Remind that there we push the calculations to YC9 lattice with width $W=9$, and the fittings are equally good, suggesting  the robustness of fittings vs. system sizes. Beyond equilibrium properties, we have also computed dynamical properties $\omega(k)$ with same parameter set on a $36\times 36$ lattice, which also show excellent agreement with experimental data (e.g., the overall dispersion line and gap values). These direct comparisons lead us unambiguously to the conclusion that the above parameters of TLI can describe the material TMGO precisely.\\

\textbf{Supplementary Note 5: The Curie-Weiss Fitting of high-$T$ susceptibility.}
\label{SI:fitchi}

As a complementary of the thermodynamic fittings in the main text, 
in Supplementary Fig.~\ref{Fig:SusHighT} we compare the experimental and simulated high-temperature magnetic susceptibility. 
It is found that the XTRG data lie on top of two experimental curves, with fitted $\Theta_{\rm{W}} \simeq 19.3$ K, 
in very good agreement with the estimates in experimental works (e.g., 18.9 - 19.1 K as indicated in the plot).
Besides, the fitted constant $C\simeq$ 210.1 cm$^3$K/mol leads to an estimate of the effective 
$g_{\parallel} = \frac{1}{\sqrt{S(S+1)} \mu_{\rm{B}}}\frac{\sqrt{3 k_{\rm{B}} C}}{\sqrt{N_{\rm{A}}\mu_0}}\simeq13.212$, 
where $S=1/2$ represents the effective spin-1/2, which is in excellent agreement with the various thermodynamic 
fittings till low temperatures in the main text and thus constitutes a self-consistency check.
Remind this $g_{\parallel}$ value obtained is in reasonably good agreement with the ideal  
Land\'e factor $g_J = 7/6$ (ideally $J_\parallel=2J g_{\rm{J}}$).

\begin{figure}[h!]
\includegraphics[angle=0,width=0.6\linewidth]{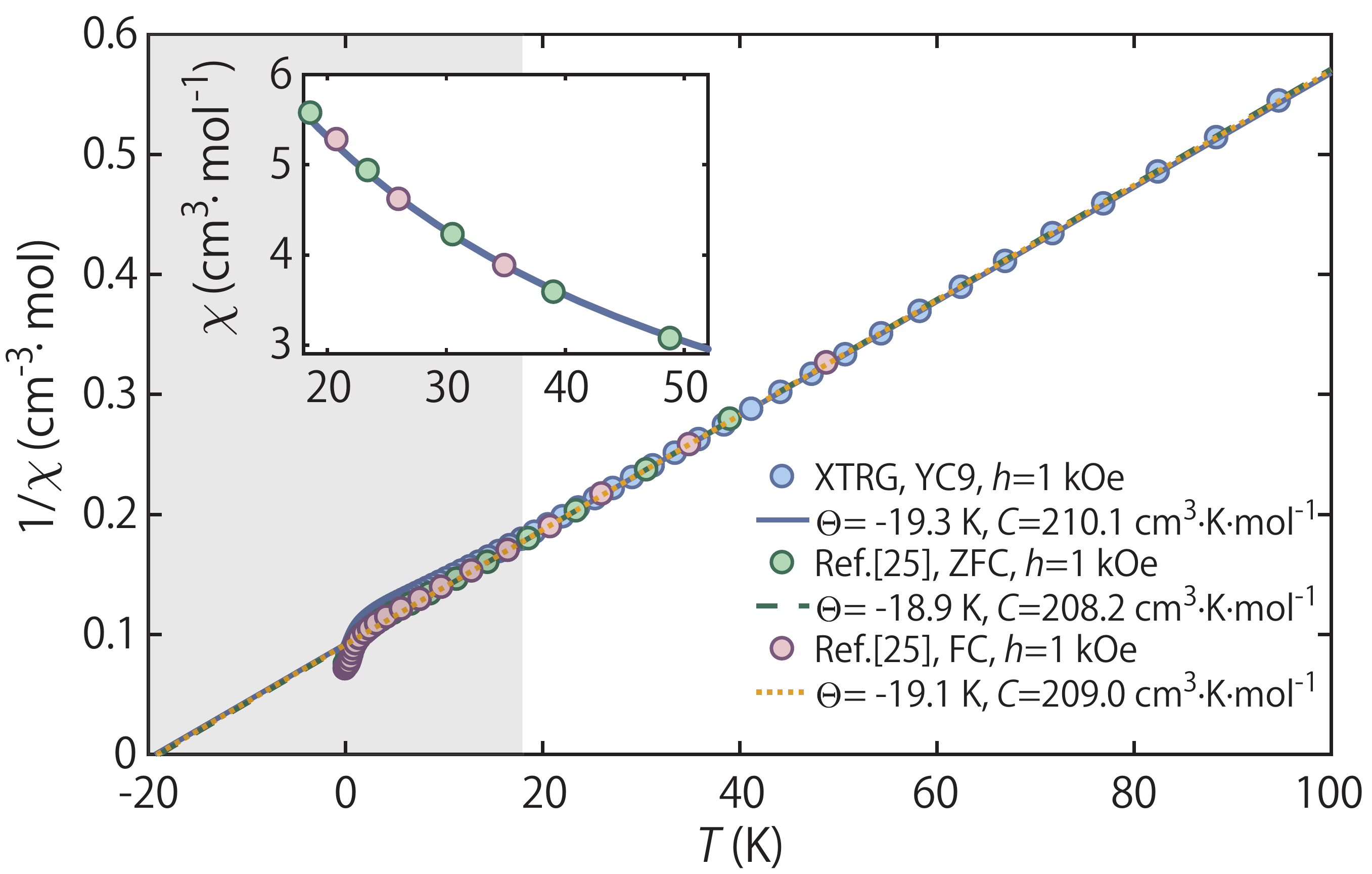}
\renewcommand{\figurename}{\textbf{Supplementary Fig.}}
\caption{\textbf{The Curie-Weiss behavior at high temperature.} 
Here we show the high-$T$ susceptibility data and analyse the Curie-Weiss behavior $\chi = \frac{C}{T-\Theta}$, 
where the inverse susceptibility $\chi^{-1}$ vs. $T$ is plotted in the main panel and 
the $\chi$ vs. $T$ curve is plotted in the inset (with Curie-Weiss fitting lines also shown). 
The XTRG susceptibility $\chi$ is computed with the model parameters $J_1=0.99$ meV, $J_2=0.05J_1$, $\Delta=0.54J_1$, 
and $h=1$ kOe, on a YC$9\times12$ lattice. 
The Curie-Weiss fittings are done within the temperature range $T\in$ [6, 48] K (experimental curves, FC and ZFC), 
and $T\in$ [18, 100] K (XTRG data), i.e., high temperature data on the right of the grey shaded regime.}
\label{Fig:SusHighT}
\end{figure}

To conclude, through large-scale simulations of both equilibrium and dynamical properties, we pinpoint the model parameters of TMGO as $J_1=0.99$ meV, $\Delta/J_1=0.54$, $J_2/J_1=0.05$, and $g_\parallel = 13.212$, which can be used to fit virtually all available experimental data, including the magnetic specific heat, entropy, susceptibility (both high- and low-temperature parts), and dynamical spin spectrum, etc.\\

\textbf{Supplementary Note 6: Specific heat curves under external fields.}

As noted in Fig.~\B{2}(d) earlier in the main text, the magnetization curves $M(h)$ show quasi-plateau structures at $M/M_{\rm{sat}} \simeq 1/3$, where $M_{\rm{sat}}$ is the saturation magnetization along the $z$ direction. In Supplementary Ref.~\cite{Li2018}, the specific heat curves have also been measured under various external magnetic fields. In Supplementary Fig.~\ref{Fig:MagC}(a), we redrawn the experimental data, and compare them, side by side, to the simulated $C_{\rm{m}}(h, T)$ curves. It can be seen that in both Supplementary Figs.~\ref{Fig:MagC}(a) and (b), a low-$T$ shoulder gradually appears under small fields, e.g., at $h=5$ kOe, and then prominent peaks show up at around $h = 10$-20 kOe. By further increasing the magnetic fields, the peak height gets declined and at the same time  its position moves towards higher temperatures for $h\geq 30$ kOe. 

\begin{figure}[h!]
\includegraphics[angle=0,width=0.8\linewidth]{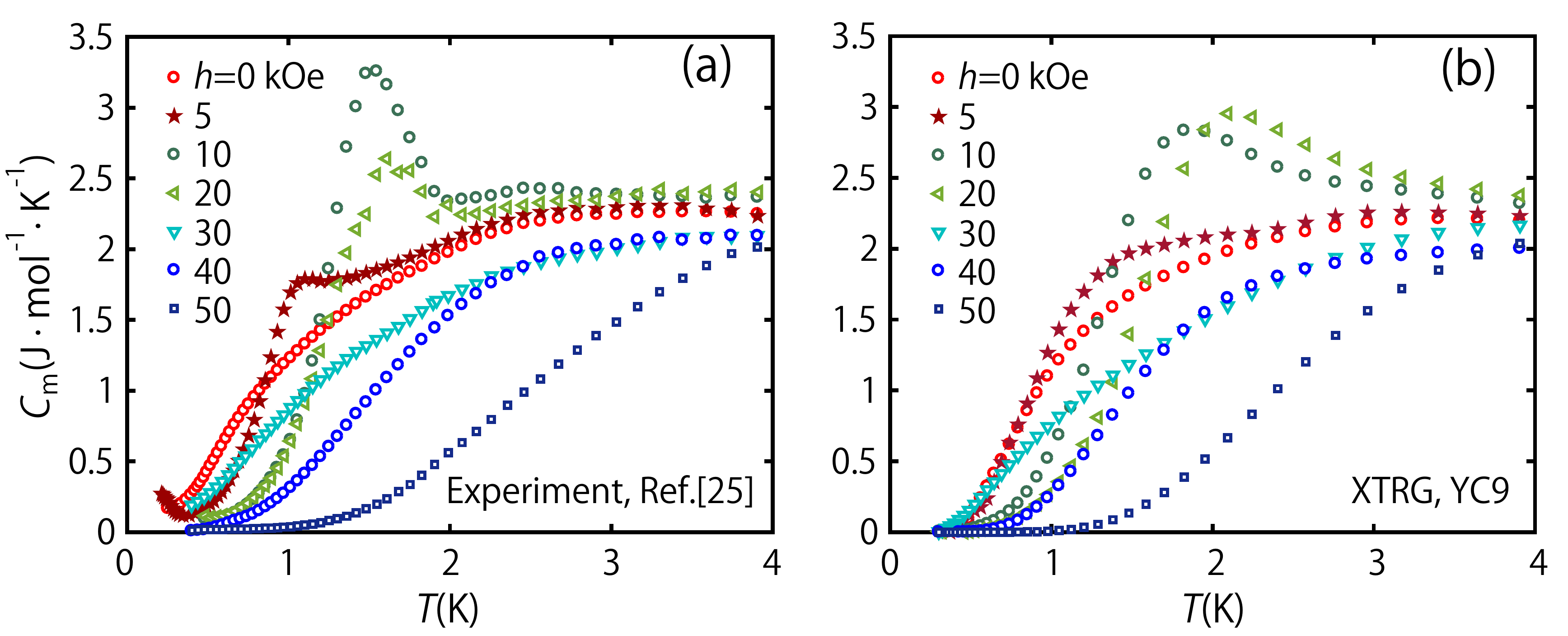}
\renewcommand{\figurename}{\textbf{Supplementary Fig.}}
\caption{\textbf{Specific heat curves under magnetic fields.} (a) shows the experimental specific heat data under various magnetic fields, taken from Supplementary Ref.~\cite{Li2018}, and (b) is the corresponding XTRG results. The comparisons are in quite remarkable consistency, suggesting the correctness of the parameter sets and the accuracy of XTRG calculations.}
\label{Fig:MagC}
\end{figure}

It is remarkable that these highly nontrivial and non-monotonic behaviors of $C_{\rm{m}}(h, T)$ curves can also be understood within the TLI model. As shown in Fig. \B{2}(d) of the main text, there exists a quasi-plateau structure at about $1/3 \, M_{\rm{sat}}$ in the curve, which suggests that the system undergoes a transition from the clock order to an UUD spin state upon increasing fields to $h=10$-20 kOe. The forming of the UUD structure releases entropy and gives rise to the prominent low-$T$ peak in $C_{\rm{m}}$, despite that the peaks in $C_{\rm{m}}$ between 1 and 2 K are less pronounced in the simulated results than experiments. As the field strength further increases, the UUD order becomes weaker and accordingly the $C_{\rm{m}}$ peak moves back to lower $T$ side with a decreasing height. Eventually, for $h\geq30$ kOe, the system gradually polarizes into a ferromagnetic spin configuration, and the hump in $C_{\rm{m}}$ moves towards higher $T$ as $h$ enhances. \\

\textbf{Supplementary Note 7: Saddle point in the triangular tight-binding model.}

When restricted in a subspace of configurations with only one pair of spins flipped (while others remain in the classical stripy order, in the small $\Delta$ limit), we consider a ``tight-binding" model 
\begin{equation}
\label{Eq:TLTB}
H = \epsilon_0 + \sum_i \sum_\delta t (c^\dagger_i c_{i+\delta} + h.c.)
\end{equation}
on the triangular lattice, where $i$ labels the lattice site, and $| i \rangle$ labels a state with a spin flipped at site $i$. $\delta = \bold{a},\bold{b},\bold{a}-\bold{b}$ denotes nearest neighboring sites ($\bold{a},\bold{b}$ are primitive vectors shown in Fig.~\B{1} of the main text). Remember that a second-order process related to $S^x$ terms in the Hamiltonian can actually tunnel between $| i \rangle$ and $| j \rangle$, so $t\sim \Delta^2$ (in the small $\Delta$ limit), given $i$ and $j$ constitute a pair of nearest neighboring sites.  

For the sake of simplicity, we set $\epsilon_0=0,t=1$, and take Fourier transformation of Eq.~(\ref{Eq:TLTB}). The resulting dispersion $\epsilon(k) = 2 [\cos(k \cdot \bold{a}) + \cos(k \cdot \bold{b}) + \cos(k \cdot (\bold{a}-\bold{b}))]$ is plotted in Supplementary Fig.~\ref{Fig:TBdisp}(a). By cutting along the $\Gamma$-M-$\Gamma$ path, we observe a quadratic low-energy dispersion near the minimum as shown in Supplementary Fig.~\ref{Fig:TBdisp}(b). On the other hand, as shown in Supplementary Fig.~\ref{Fig:TBdisp}(c) the M point constitutes a maximal along K-M-K path. Therefore, it is evident that the M point indeed constitutes a saddle point in the dispersion. \\

\begin{figure}[h!]
\includegraphics[angle=0,width=0.75\linewidth]{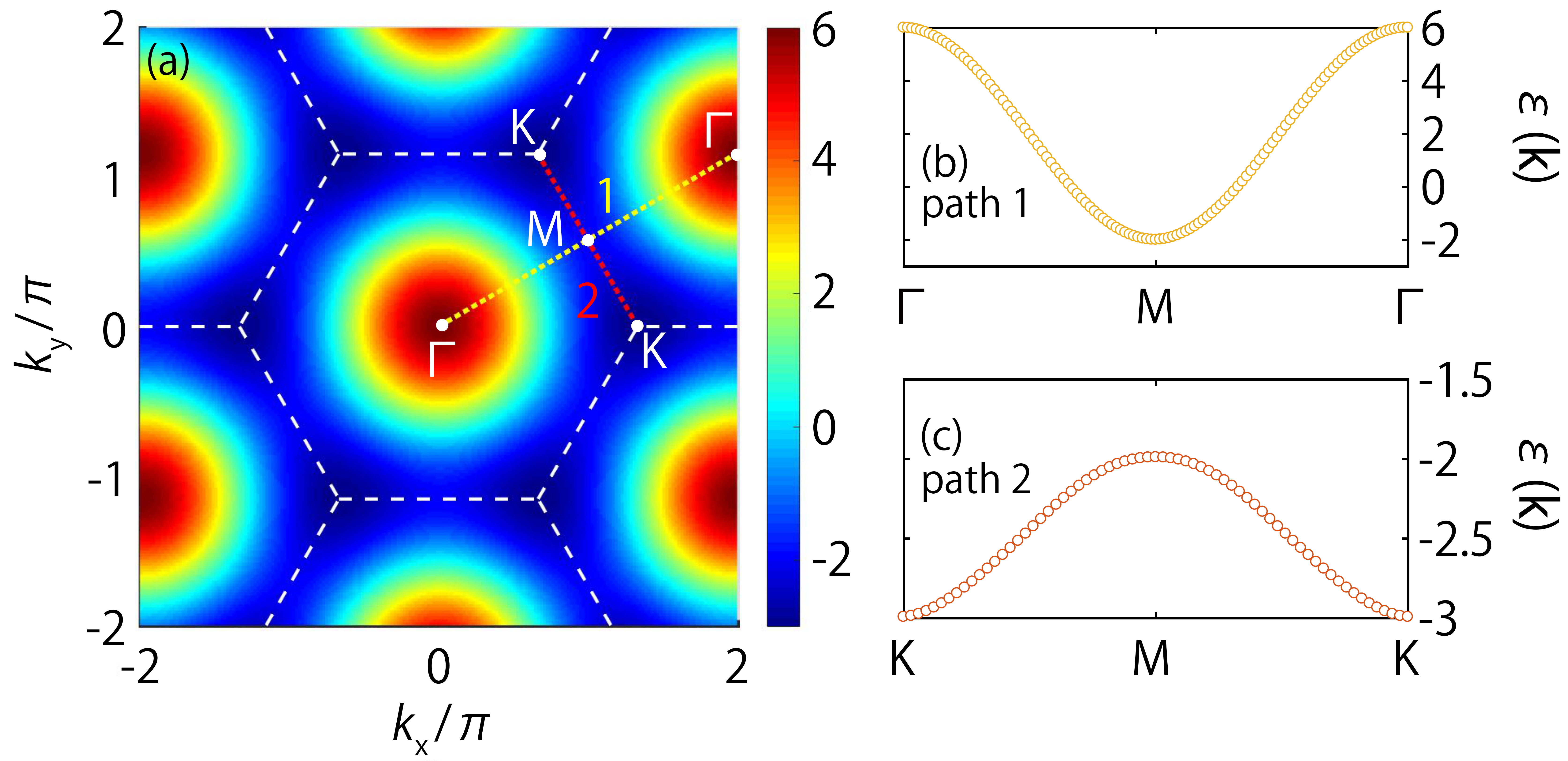}
\renewcommand{\figurename}{\textbf{Supplementary Fig.}}
\caption{\textbf{``Tight-binding" dispersion on triangular lattice and the saddle point at M.} 
(a) shows the contour plot of dispersion $\epsilon(\bf{k})$, and (b,c) show the cuts along paths 1 and 2, respectively.}
\label{Fig:TBdisp}
\end{figure}

\textbf{Supplementary Note 8: A brief introduction of path integral QMC for TLI.}

The Hamiltonian of quantum TLI is
$\mathcal{H} = J_1\sum_{\langle i,j \rangle} S_i^z S_j^z + J_2 \sum_{\langle\langle i,j \rangle\rangle} S_i^z S_j^z - h\sum_i S_i^x $,
where the external magnetic field part is omitted for clearance of narrative. 
QMC evaluates the partition function $Z=\text{Tr}\left[e^{-\beta\mathcal{H}}\right]$, 
where discretized imaginary time is used and $\beta=L_\tau\Delta \tau $ ($\Delta \tau=0.05$). 
It maps the problem onto a (2+1)D classical system as follows,
\begin{equation}
\begin{split}
Z & =  \text{Tr}\left[e^{-\beta\mathcal{H}}\right]\\
& =  \text{Tr}\left[e^{-\Delta\tau\mathcal{H}}e^{-\Delta\tau\mathcal{H}} \cdots e^{-\Delta\tau\mathcal{H}}e^{-\Delta\tau\mathcal{H}}\right].
\end{split}
\end{equation}
By inserting a complete set of $S^z_{\rm{i}}$ eigenstates between each pair of exponentials, i.e.,
\begin{equation}
\begin{split}
	\mathbf{1} &=\prod_{i=1}^N\left[ \sum_{S_i^z=\pm 1} \vert S_i^z\rangle\langle S_i^z\vert \right]\\
	&\equiv \sum_{\{S_i^z\}}\vert S^z\rangle\langle S^z\vert,
\end{split}
\end{equation}
we can then rewrite the partition function as,
\begin{equation}
\begin{split}
Z=\sum_{\{S_{i,l}=\pm 1\}} & \langle S_1^z\vert e^{-\Delta\tau \mathcal{H}} \vert S_L^z\rangle \langle S_L^z\vert e^{-\Delta\tau \mathcal{H}} \vert S_{L-1}^z\rangle \langle S_{L-1}^z\vert e^{-\Delta\tau \mathcal{H}} \vert S_{L-2}^z\rangle\cdots  \\
  & \cdots \langle S_3^z\vert e^{-\Delta\tau \mathcal{H}} \vert S_2^z\rangle \langle S_2^z\vert e^{-\Delta\tau \mathcal{H}} \vert S_1^z\rangle,
\end{split}
\end{equation}
where $l$ indices the time slice $\tau=l\cdot\Delta\tau$. 

Next, we employ the Trotter-Suzuki decomposition,
\begin{equation}
\begin{split}
\langle S_{l+1}^z\vert e^{-\Delta\tau \mathcal{H}} \vert S_l^z\rangle &= \langle S_{l+1}^z\vert e^{-\Delta\tau \mathcal{H}_1-\Delta\tau \mathcal{H}_0} \vert S_l^z\rangle\\
&= \langle S_{l+1}^z\vert e^{-\Delta\tau \mathcal{H}_1}e^{-\Delta\tau \mathcal{H}_0} \vert S_l^z\rangle +\bm{O}[(\Delta\tau)^2],
\end{split}
\end{equation}
where $\mathcal{H}_0=J_1\sum_{\langle i,j \rangle} S_i^z S_j^z + J_2 \sum_{\langle\langle i,j \rangle\rangle} S_i^z S_j^z$ 
and $\mathcal{H}_1=- h\sum_i S_i^x $. 
Therefore the corresponding matrix element reads
\begin{equation}
\langle S_{l+1}^z\vert e^{-\Delta\tau \mathcal{H}_1}e^{-\Delta\tau \mathcal{H}_0}\vert S_l^z\rangle = \Lambda^N e^{-\Delta\tau J_1\sum_{\langle i,j \rangle} S_{i,l}^z S_{j,l}^z-\Delta\tau J_2\sum_{\langle\langle i,j \rangle\rangle} S_{i,l}^z S_{j,l}^z +\gamma\sum_{i}S_{i,l}^z S_{i,l+1}^z},
\end{equation}
with $\gamma=-\frac{1}{2}\ln\tanh(\Delta\tau h)$, and $\Lambda^2=\sinh(\Delta\tau h)\cosh(\Delta\tau h)$. For a certain configuration $\{S^z_{i,l}\}$, the configurational weight is,
\begin{equation}
\omega\{S^z_{i,l}\}=\left( \prod_l \prod_{\langle i,j \rangle}e^{-\Delta\tau J_1 S_{i,l}^z S_{j,l}^z} \right)\left( \prod_l \prod_{\langle\langle  i,j \rangle\rangle}e^{-\Delta\tau J_2 S_{i,l}^z S_{j,l}^z} \right)\left( \prod_\tau \prod_{\langle l,l' \rangle}\Lambda e^{\gamma S_{i,l}^z S_{i,l'}^z} \right) .
\end{equation}
Now, the 2D quantum problem becomes a (2+1)D classical Ising model, which can be solved by local or global update schemes,  both adopted in our practical Monte Carlo samplings. 

Within this framework, the physical observables can be evaluated as
\begin{equation}
\langle \hat{O}\rangle \approx \frac{1}{N}\sum_{p}^{N}\hat{O}(\{S_i\}_p)
\end{equation}
where $\{S_i\}_p$ denotes the spin configurations in which the measurement is performed at time $p$ of the Markov chain. Besides, we are also interested in the imaginary time spin-spin correlation function,
\begin{equation}
G(\mathbf{q},\tau) = \frac{1}{N} \sum_{i,j} e^{i\mathbf{q}\cdot\mathbf{r}_{ij}}\langle S^{z}_{i}(\tau)S^{z}_{j}(0)\rangle,
\end{equation} 
which should be calculated in prior to the spin spectrum $S(\mathbf{q},\omega)$. The latter can be obtained after a stochastic analytical continuation, as detailed in Methods section of the main text.

\end{document}